\documentclass[sigconf]{acmart}

\usepackage{subcaption}
\usepackage{graphicx}

\usepackage{fancyhdr}

\usepackage{adjustbox}
\usepackage{longtable}
\usepackage{multirow}
\usepackage{wrapfig}
\usepackage{url}
\usepackage{amsmath,amsthm}

\DeclareMathOperator*{\E}{\mathbb{E}}

\usepackage{balance}
\AtBeginDocument{%
  \providecommand\BibTeX{{%
    \normalfont B\kern-0.5em{\scshape i\kern-0.25em b}\kern-0.8em\TeX}}}

\copyrightyear{2020}
\acmYear{2020}
\setcopyright{acmcopyright}\acmConference[KDD '20]{Proceedings of the 26th ACM SIGKDD Conference on Knowledge Discovery and Data Mining}{August 23--27, 2020}{Virtual Event, CA, USA}
\acmBooktitle{Proceedings of the 26th ACM SIGKDD Conference on Knowledge Discovery and Data Mining (KDD '20), August 23--27, 2020, Virtual Event, CA, USA}
\acmPrice{15.00}
\acmDOI{10.1145/3394486.3403262}
\acmISBN{978-1-4503-7998-4/20/08}

\begin{document}
\fancyhead{} % remove head and foot

%%
%% The "title" command has an optional parameter,
%% allowing the author to define a "short title" to be used in page headers.
\title{On Sampling Top-K Recommendation Evaluation}

%%
%% The "author" command and its associated commands are used to define
%% the authors and their affiliations.
%% Of note is the shared affiliation of the first two authors, and the
%% "authornote" and "authornotemark" commands
%% used to denote shared contribution to the research.
\author{Dong Li}
\email{dli12@kent.edu}
\affiliation{%
  \institution{Kent State University}
}

\author{Ruoming Jin}
\email{rjin1@kent.edu}
\affiliation{%
  \institution{Kent State University}
}

\author{Jing Gao}
\email{jgao@ilambda.com}
\affiliation{%
  \institution{iLambda}
}

\author{Zhi Liu}
\email{zliu@ilambda.com}
\affiliation{%
  \institution{iLambda}
}

%%
%% By default, the full list of authors will be used in the page
%% headers. Often, this list is too long, and will overlap
%% other information printed in the page headers. This command allows
%% the author to define a more concise list
%% of authors' names for this purpose.
%%\renewcommand{\shortauthors}{Trovato and Tobin, et al.}

%%
%% The abstract is a short summary of the work to be presented in the
%% article.
\begin{abstract}
Recently, Rendle has warned that the use of sampling-based top-$k$ metrics might not suffice. This throws a number of recent studies on deep learning-based recommendation algorithms, and classic non-deep-learning algorithms using such a metric, into jeopardy. In this work, we thoroughly investigate the relationship between the sampling and global top-$K$ Hit-Ratio (HR, or Recall), originally proposed by  Koren~\cite{CremonesiKT@10} and extensively used by others. 
By formulating the problem of aligning sampling top-$k$ ($SHR@k$) and global top-$K$ ($HR@K$) Hit-Ratios through a mapping function $f$, so that $SHR@k\approx HR@f(k)$, we demonstrate both theoretically and experimentally that the sampling top-$k$ Hit-Ratio provides an accurate approximation of its global (exact) counterpart, and can consistently predict the correct winners (the same as indicate by their corresponding global Hit-Ratios).  
\end{abstract}

%%
%% The code below is generated by the tool at http://dl.acm.org/ccs.cfm.
%% Please copy and paste the code instead of the example below.
%%



%%
%% Keywords. The author(s) should pick words that accurately describe
%% the work being presented. Separate the keywords with commas.

%% A "teaser" image appears between the author and affiliation
%% information and the body of the document, and typically spans the
%% page.

%%
%% This command processes the author and affiliation and title
%% information and builds the first part of the formatted document.
\maketitle

\section{Introduction}

% Recommendation systems (RS) have become a vital and integral part of the modern web, enabling the key services of both internet giants, such as Amazon, Netflix, Alibaba, etc., and millions of smaller e-commerce and online media websites. In order to better serve their customers, and to gain and maintain their competitive advantages, these companies need to continue to invest in and to improve their recommendation systems. The importance of recommendation systems is found in the core premise of the internet: connecting users with the right products, information, and/or other users. When a website offers tens of thousands to multiple billions of products, it simply goes beyond an individual customer's capability to browse and search. Thus, recommendation, or personalization, which aims to best match the preferences and/or needs of an individual customer across all the available choices, is simply indispensable. Concretely, a recommendation system computes a list of ``personalized'' items pertaining to each individual user, based on her or his profile and past behaviors, such as view/click histories, etc. The typical strategy of recommendation ranks all available items, according to certain relevance scores, based on collaborative filtering and contents, and selects the top-$k$ items. 

Over the last few years, in both industry and academic research communities, many efforts have been taken to integrate deep learning into recommendation techniques~\cite{zhang2017deep}.  Though the flourishing list of publications has demonstrated sizeable improvements over the classical non-deep (linear) approaches, several recent studies ~\cite{RecSys19Evaluation, Steffen@19DBLP} have sounded the alarm: The displayed success in recommendation may contribute to the weaker baseline. Some other factors, such as evaluation protocols and performance measures, together with choices of datasets, may also play roles in the potentially over-promising results ~\cite{RecSys19Evaluation}. 

Lately, Rendle ~\cite{rendle2019evaluation} has noticed that, in recent deep learning-based recommendation studies, it is becoming popular ~\cite{he2017neural, ebesu2018collaborative,HuSWY18,krichene2018efficient,wang2018explainable,YangBGHE18,YangCXWB18} 
to adopt sampling-based criteria for top-$k$ evaluation. Basically, instead of ranking all available items, which might be a very large list, these studies sample a smaller set of (irrelevant) items, and rank the relevant items against the sampled items. In his findings, he claimed that the typical process used top-$k$ evaluation metrics, such as Recall/Precision (Hit-Ratio), Average Precision (AP) and nDCG, other than (AUC), are all ``inconsistent'' with respect to the exact metrics (even in expectation). He even suggests avoiding using the sampled metrics for top-$k$ evaluation. 

What does this mean to all the existing studies which use sampling top-$k$ criteria? Have their results all become (sort-of) invalid? Does that mean we have to use all the items for any meaningful top-$k$ evaluation? Clearly, a significant amount of efforts in the recommendation community is at risk here.  To be able to firmly answer these questions, a better understanding of the sampling-based top-$k$ metrics is much needed. In the meantime, a sampling approach, where acceptable, can be a useful tool for saving computational cost and speeding up evaluation time. While computational resources might not be a big problem for enormous mega-corporations, such as Google or Amazon, for many smaller, resource-constrained organizations, it may still be an issue. For instance, if valid, a sampling approach can be a quick way to help evaluate the promise of a given algorithm, screening for the eventual exact/global top-$k$ evaluation. 

\noindent{\bf A Bit of History on Sampling Top-$k$ Evaluation:} 
The sampling top-$k$ method was initially suggested by Koren in the seminal work~\cite{Koren08} as an approach to measure the success of top-$k$ recommenders. Specifically, he uses $1000$ additional random movies (which may include already-ranked ones) against the targeted movie $i$ for a user. He ranks these $1001$ movies by the predicted rating (relevance score), and he normalizes the ranking score between $0$ and $1$. Finally, he draws the cumulative distributions of all the users, with respect to the ranking score. Essentially, for any algorithm, at a given rank $k$, the value (the point in the cumulative distribution curve) is basically Hit-Ratio or Recall (at $k$) under sampling. Another highly cited work~\cite{CremonesiKT@10} has utilized this metric to evaluate the performance of a variety of recommendation algorithms on Top-$N$ recommendation tasks. 

This method was first adopted by deep learning-based recommendation papers in ~\cite{ElkahkySH@WWW'15} and then in ~\cite{he2017neural}. Here, the authors go beyond the top-$k$ Hit-Ratio suggested by Koren~\cite{he2017neural,ElkahkySH@WWW'15}, extending to metrics such as Mean Reciprocal Rank (MRR) and nDCG. Since Koren, various other deep learning-based recommendation studies ~\cite{ebesu2018collaborative,HuSWY18,krichene2018efficient,wang2018explainable,YangBGHE18,YangCXWB18} have adopted such sampling-based top- $k$ evaluation metrics. In these studies, they typically sample only those ``irrelevant'' items (not scored by the users), unlike the work in Koren, which may sample relevant items, as well. The number of items sampled typically ranges from $100$ to $1000$. 

However, besides the latest study and warning by Rendle~\cite{rendle2019evaluation}, there have been no studies on the statistical properties of sampling-based top-$k$ evaluation metrics. Clearly, as indicated by Rendle~\cite{rendle2019evaluation}, the sampling top-$k$ metric is very different from global top-$k$ metric. But do they relate to each other? Can sampling top-$k$ reflect global (exact) top metrics? And how do we interpret the existing experimental results which use sampling-based metrics? 

\noindent{\bf Our Contributions:}
To answer the above questions, we perform the first study to thoroughly investigate the relationship between the sampling and global top-$k$ Hit-Ratio (HR, or Recall), originally proposed by Koren~\cite{Koren08}. Top-$k$ hit ratio is one of the most popular metrics used for evaluating almost all top-$N$ recommenders~\cite{zhang2017deep}. Specifically, we made the following contribution: 

%\begin{itemize}[leftmargin=*,noitemsep,nolistsep]
\begin{itemize}

\item (Section~\ref{problem}) We formalize the problem of aligning sampling top-$k$ ($SHR@k$) and global top-$K$ ($HR@K$) Hit-Ratios through a mapping function, so that $SHR@k\approx HR@f(k)$, where $f$ is the functions map of the $k$ in the sampling to global top $f(k)$. We also prove the {\em Sampling Theorem}, which shows the sampling Hit-Ratio preserves the ``dominating'' property between global Hit-Ratios. 
%In this way, we aim to show that the sampling metric is similar to the ``signal sampling''~\cite{rao2018signals}, where the global metrics curve between top $1$ to $N$ (the entire population of items) is sampled (and approximated) at $f(1)<f(2)<..<f(n)$ locations, which corresponds to the $SHR@K$ curve ($K=1,2, \cdots, n$). 
\item (Sections~\ref{estimator} and ~\ref{specific}) We develop novel methods to approximate function $f$, and we show that it is surprisingly approximately linear, even under non-linear computation (when $k$ is large). Basically, the ``sampling'' location of the global top-$K$ curve is almost equally intervaled ($f(i)-f(i-1)$ is close to constant).  In addition, we develop algorithm-specific mapping functions and discuss a list of key properties to help ensure the predictive power of sampling. 
    % In addition, $SHR@K$ is closely centered around its expectations for large populations (thus can avoid the need for multiple samples). 
    \item (Section~\ref{experiments})  We experimentally validate our mapping function $f$ by comparing between the sampling and its global top-$K$ counterparts, and we show that our function can provide a rather accurate estimate of its global top-$f(k)$. We also show that the sampling Hit-Ratio can accurately predict the same winners as the corresponding global Hit-Ratio.
 \end{itemize}
 
    %\item (Section~\ref{rerank}) Most of the existing studies focus on rather small $k$. By comparing different algorithms on the entire cumulative distribution (curve), using sampling and global top-$k$ Hit-Ratio, we that observe different algorithms may perform better at different stages: Some are better on the smaller $k$, while others are better on larger $k$. Given this, we found a simple re-ranking strategy: Using the former to rank the latter's top-$k$ list can improve the overall top-$k$ Hit-Ratio curve (as well as other measures, such as nDCG). This provides a theoretical rationale as to why industry practitioners' often use two recommendation algorithms in a recommendation system: one in the first stage, to provide a candidate list (improving recall); the other, to further rank the candidate list for the final top-$k$ computation.  

\section{Problem and Sampling Theorem}
\label{problem}
%Dong, add subsection

\begin{scriptsize}
      \begin{table}
      \begin{minipage}{0.48\textwidth}
      \caption{Notations}
      \vspace{-2.0ex}
  \label{tab:notation}
 \begin{tabular}{|l|l|}
 \hline
    $I$	& entire set of items, and its size $|I| = N$\\
    \hline
    $i_u$	& relevant item for user $u$ in testing data\\
    \hline
    $I_n$	& sampled item set (user-specific), composed of $n-1$ sampled items and $i_u$\\
    \hline
    $R$	& integer variable, referring to item rank position, in range $[1, N]$\\
    \hline
	$R_u$	& rank of item $i_u$ among $I$ for user $u$	\\
	\hline
	$r^u$	& rank of item $i_u$ among $I_n$ for user $u$; also denotes a random variable\\
	\hline
	$r^R$	& $:= r^u$, for the group of users whose $R_u = R$ \\
	\hline
	$HR@K$	& global  top-K hit-ratio (recall), Formula~\ref{eq:HRK_def} \\
	\hline
	$SHR@k$	& sampling top-k hit-ratio (recall), Formula~\ref{eq:SHR_def0}\\
    \hline
    $f$& mapping $SHR@k$ to $HR@K$, where $K = f(k)$\\
    \hline
    $W_R$ & fraction of users where $R_u = R$, (Eq ~\ref{eq: PR_def}); also denotes user ranking distribution\\
    \hline
    $p^u$ & $Pr(r^u\le k)$, probability that $r^u \le k$ among $I_n$ for a user $u$\\
    \hline
    $p^R$ & $:= p^u$, for the group of users whose $R_u = R$\\
    \hline
    $p_u$&$:= \frac{R_u - 1}{N-1}$, probability for sampling an item that ranks higher than $i_u$ for $u$\\
    \hline
  \end{tabular}
  \end{minipage}
  \end{table}
\end{scriptsize}

\subsection{Problem Formulation}
 Assume we split the entire dataset into training and testing. Let the testing dataset consist of $M$ users and $N=|I|$ items, where $I$ is the entire set of items. From the training, we will learn a recommendation algorithm $A$, which can rank a given set of items for a user.  To simplify our discussion, we consider the \textit{leave-one-out} strategy ~\cite{mmanuelXBS@16}, where {\bf each user has and only has one relevant item to be evaluated}, though such treatment can be naturally generalized to the situation where a user may have more than one targeted item in the testing data~\cite{ElkahkySH@WWW'15,liang2018variational}.  Table ~\ref{tab:notation} highlights the key notations used in the rest of the paper.

%Dong, change to subsubsection
\subsubsection{Global Top-K Hit Ratio}
Given a user $u$, and its relevant item $i_u$ $(i_u\in I)$ in the testing dataset, the recommendation algorithm $A$ will calculate the relative rank of the relevant item $i_u$, denoted as $R_u$, among all available items $I$, $R_u=A(u, i_u, I)$.

Let us consider the global top-$K$ Hit-Ratio (or recall) metric: 
%Dong, change P_R to W_R
% \begin{small}
% \begin{equation}\label{eq:HRK_def}
% \begin{split}
%     HR@K = \frac{1}{M}\sum\limits_{u = 1}^{M}{{\bf 1}_{R_u \le K}} = \sum\limits_{R = 1}^{N}{P_R \cdot {\bf 1}_{R\le K}}
%     \end{split}
% \end{equation}
% \end{small}

\begin{small}
\begin{equation}\label{eq:HRK_def}
\begin{split}
    HR@K = \frac{1}{M}\sum\limits_{u = 1}^{M}{{\bf 1}_{R_u \le K}} = \sum\limits_{R = 1}^{N}{W_R \cdot {\bf 1}_{R\le K}}
    \end{split}
\end{equation}
\end{small}

%Dong, change P_R to W_R
% Here, ${\bf 1}_X$ is the indicator function of event $X$ (${\bf 1}_X=1$ iff $X$ is true and $0$ others), and $P_R$ is the frequency of users with rank $R$:  
% \begin{small}
% \begin{equation}\label{eq: PR_def}
%     P_R = \frac{1}{M}\sum_{u = 1}^{M}{{\bf 1}_{R_u = R}}
% \end{equation}
% \end{small}

Here, ${\bf 1}_X$ is the indicator function of event $X$ (${\bf 1}_X=1$ iff $X$ is true and $0$ others), and $W_R$ is the frequency of users with item $i_u$ rank in position $R$:  
\begin{small}
\begin{equation}\label{eq: PR_def}
    W_R = \frac{1}{M}\sum_{u = 1}^{M}{{\bf 1}_{R_u = R}}
\end{equation}
\end{small}

% we take $K$ to stress that is ranked in population $I_N$ and $k$ in the sample set $I_n$. The HR at $top@k$ evaluated among population is $HR@K$ and $SHR@k$ is the one among sample set. 
% Note, $HR@K$ and $SHR@k$ are discrete points , all possible values form discrete functions $HR(K)$ and $SHR(k)$.Naturally, they could be generated to smooth function in real range when necessary. For example, $HR(m + 0.5) = \frac{HR(m) + HR(m + 1)}{2}$ for integer $m$ . 

%The $HR@K$ is defined as:

%The performance of the algorithm will be evaluated by metrics typical some $top@k$ methodology, like hit ratio(HR), $Recall@k$, $Precision$@k, ndcg, etc,  based on the relative rank of target items.

% In this paper, we evaluate the algorithms under $HR@k$ metric and study the performance relationship in between the whole available item set and the sampled items.

%  Now there are two ways to evaluate the performance of an algorithm. One is, for each user, compare the relevance of all $N$ user item pairs and get the rank of target item, noted as $R_u = A(u, I_N)$, where $I_N$ is the whole item set. The other is, 

%Dong, change to subsubsection
\subsubsection{Sampling Top-k Hit Ratio}
Next, let us revisit the top-$k$ Hit-Ratio under sampling. For a given user $u$ and the relevant item $i_u$, we first sample $n - 1$ items from the entire set of items $I$, forming the subset $I_n$ (including $i_u$). Let the relative rank of $i_u$ among $I_n$ be denoted as $r^u = A(u, i_u, I_n)$. Note that $r^u$ is a random variable depending on the sampling set $I_n$. 

Given this, the sampling top-$k$ Hit-Ratio can be defined as $SHR@k$: 
\begin{small}
\begin{equation}\label{eq:SHR_def0}
    SHR@k=\frac{1}{M} \sum\limits_{u=1}^{M}{Z^u}, \ \ Z^u \sim Bernoulli(p^u=Pr(r^u\le k))
\end{equation}
\end{small}
where, $Z^u$ is a random variable for each user $u$, and follows a Bernoulli distribution with probability $p^u=Pr(r^u\le k)$. 

%Dong, add one sentence
Now, recall that we are trying to study the relation between $SHR@k$ and $HR@K$.

%Dong, add subsection
\subsection{Sampling}
We note that the population sum $\sum_{u=1}^M Z^u$ is a Poisson binomial distributed variable (a sum of $M$ independent Bernoulli distributed variables). Its mean and variance will simply be sums of the mean and variance of the $n$ Bernoulli distributions:
\begin{small}
\begin{equation*}
    \mu = \sum_{u=1}^M{p^u}, \ \ \ \ 
    \sigma^2 = \sum_{u=1}^M p^u (1-p^u)
\end{equation*}
\end{small}

Given this, the expectation and variance of $SHR@k$:
%Dong, change P_R to W_R
% \begin{small}
% \begin{equation}\label{eq: ESHR_def0}
% \begin{split}
%     &E[SHR@k] = \frac{1}{M} \sum\limits_{u=1}^{M}{p^u} 
%     = \sum\limits_{R = 1}^{N}{P_R \cdot p^R}  
% \end{split}
% \end{equation}
% \begin{equation}\label{eq: ESHR_var}
% \begin{split}
%     &Var[SHR@k] = \frac{1}{M^2} 
%     \sum\limits_{u=1}^{M}{p^u(1-p^u)} 
%     =\frac{1}{M} \sum\limits_{R = 1}^{N}{P_R \cdot p^R(1-p^R)}  
% \end{split}
% \end{equation}
% \end{small}

\begin{small}
\begin{equation}\label{eq: ESHR_def0}
\begin{split}
    &E[SHR@k] = \frac{1}{M} \sum\limits_{u=1}^{M}{p^u} 
    = \sum\limits_{R = 1}^{N}{W_R \cdot p^R}  
\end{split}
\end{equation}
\begin{equation}\label{eq: ESHR_var}
\begin{split}
    &Var[SHR@k] = \frac{1}{M^2} 
    \sum\limits_{u=1}^{M}{p^u(1-p^u)} 
    =\frac{1}{M} \sum\limits_{R = 1}^{N}{W_R \cdot p^R(1-p^R)}  
\end{split}
\end{equation}
\end{small}
%Dong, change P_R to W_R; modify the explanation
% where $p^R$ is any user in the group where $R_u = R$, and they all have the same $p^u$,  and $W_R$ is defined in equation \ref{eq: PR_def}.
 The probability for users who are in the same group ($R_u = R$), share the same $p^u$, will be denoted by $p^R$  and $W_R$ is defined in equation \ref{eq: PR_def}.

To define $p^u=Pr(r^u \le k)$ precisely, let us consider the two commonly used types of sampling (with and without replacements). 

\noindent{\bf Sampling with replacement (Binomial Distribution):}

% Immediately, we can derive the equation of HR in sample set:
% \begin{equation}\label{eq:SHR1}
%     SHR@k = \frac{1}{M} \sum\limits_{u = 1}^{M}{\delta(r_u \le k)}
% \end{equation}

% We start from sampling procedure to derive the relationship of the HR metric between sets.

% A specific user $u$ ($R_u$ naturally given), $n-1$ items will be sampled from the $N-1$ items either with or without replacement.

% \subsubsection{Sampling with replacement: Binomial Distribution}

For a given user $u$, let $X^u$ denote the number of sampled items that are ranked in front of relevant item $i_u$: 
\begin{small}
\begin{equation*}\label{eq:bern_dist}
    X^u = \sum\limits_{i=1}^{n-1}{X_i^u}, \quad X_i^u \sim Bernoulli(p_u = \frac{R_u - 1}{N-1}) 
\end{equation*}
\end{small}
where  $X_i^u$ is a Bernoulli random variable for each sampled item $i$: $X_i^u=1$ if item $i$ has rank range in $[1, R_u -1]$ ($p_u$ is the corresponding probability) and $X_i^u=0$ if $i$ is located in $[R_u + 1, N]$. Thus, $X^u$ follows binomial distribution: 
\begin{small}
\begin{equation}\label{eq:binom_dist}
\begin{split}
    \quad X^u \sim Binomial(n-1, p_u=\frac{R_u - 1}{N-1}) 
    \end{split}
\end{equation}
\end{small}
And the random variable $r^u=X^u +1$, and we have 
\begin{small}
\begin{equation*}
\begin{split}
    p^u & =CDF(k;n-1,p_u)=Pr(r^u\le k) \\
    & = \begin{cases}
    \sum\limits_{l=0}^{k-1}{    {\binom{n-1}{l}} p_{u}^{l} (1-p_{u})^{n-1-l}    } &, R_u \geq k\\
    1 &, R_u < k
    \end{cases}
    \end{split}
\end{equation*}
\end{small}
\noindent{\bf Sampling without replacement (Hypergeometric Distribution):} 
If we sample $n-1$ items from the total $N-1$ items without replacement, and the total number of successful cases is $R_u-1$, then let $X^u$ be the random variable for the number of items appearing in front of relevant item $i_u$ ($r^u=X^u+1$):
\begin{small}
\begin{equation*}
    X^u\sim Hypergeometric(N-1, R_u-1, n-1)
\end{equation*}
 \begin{equation*}
 \begin{split}
    p^u & =CDF(k;N-1, R_u-1, n-1)=Pr(r^u \le k) \\
    & =\begin{cases}
    \sum\limits_{l=0}^{k-1}{\frac{{\binom{R_u-1}{l}}{\binom{N - R_u}{n - 1 - l}} }{{\binom{N-1}{n -1}} }} &, R_u \geq k\\
    1 &, R_u < k
    \end{cases}
    \end{split}
\end{equation*}
\end{small}

%Dong, add detailed caption
\begin{small}
\begin{figure*}[th]
\begin{subfigure}{0.45\textwidth}
    \includegraphics[width=\textwidth, height = \textwidth]{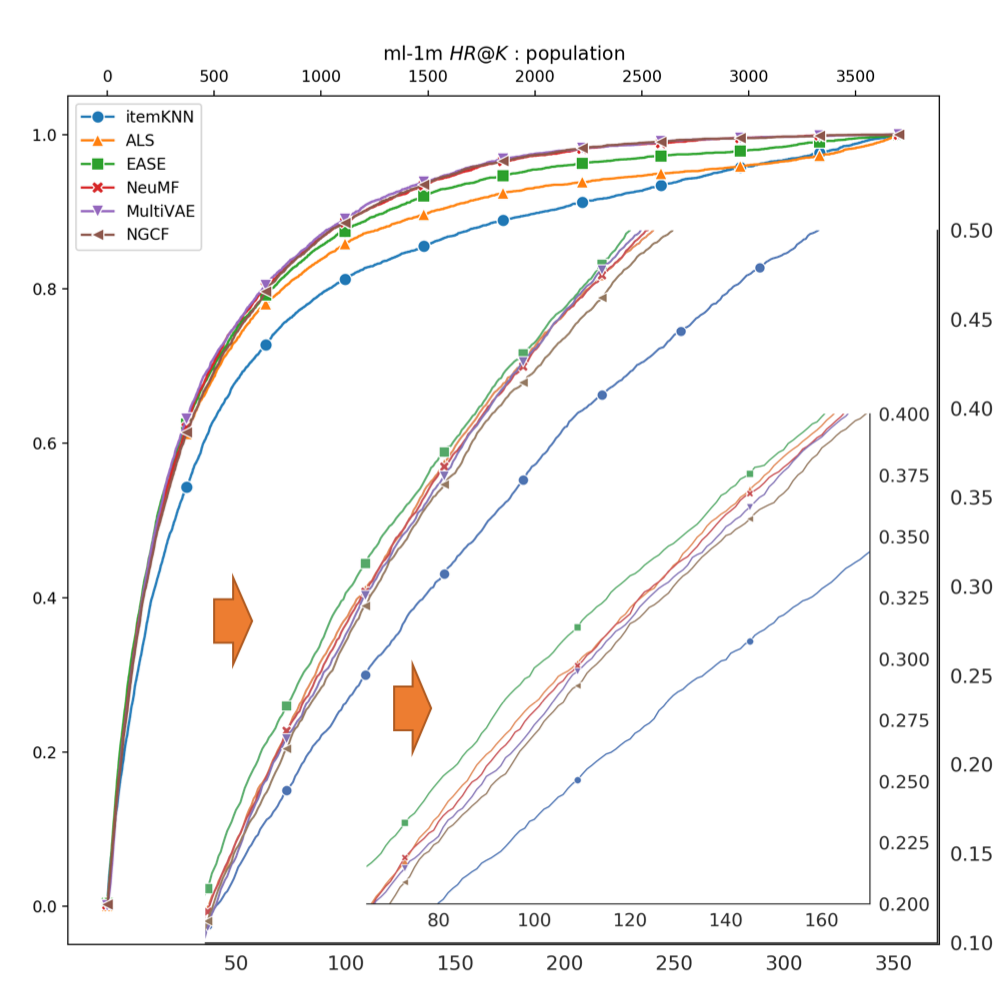}
    \caption{Global Top-$K$, $HR@K$}
    \label{fig:1a}
\end{subfigure}
\begin{subfigure}{0.45\textwidth}
    \includegraphics[width=\textwidth,height = \textwidth]{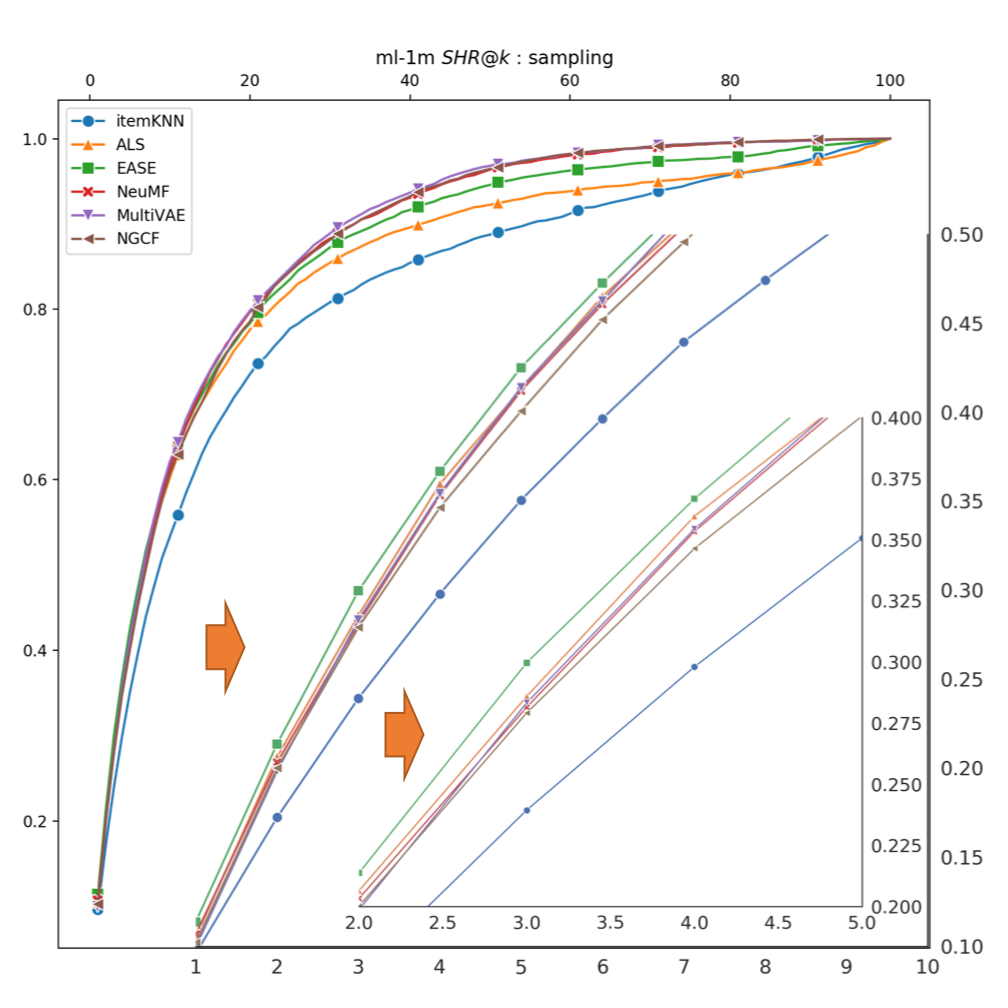}
    \caption{Sampling Top-$K$, $SHR@k$}
    \label{fig:1b}
\end{subfigure}
\caption{Global vs Sampling Top $k$ Hit-Ratio on MovieLens 1M dataset (ml-1m).
To display the details clearly, we zoom in at different range scales. Compare two figures, we can easily conclude that sampling evaluation maintains the same trend as global evaluation for different algorithms even at small error range.}
\label{figure:ml-1m:correspondence}

\end{figure*}
\end{small}
It is well-known that, under certain conditions, the hypergeometric distribution can be approximated by binomial distribution. We will focus on using binomial distribution for analysis, and we will validate the results on  hypergeometric distribution experimentally. 

% \noindent{\bf Poisson Binomial Distribution:}
%\vspace*{-1.0ex}
\subsection{A Functional View of $HR@K$ and $SHR@k$}
To better understand the relationship between $HR@K$ (global top-K Hit-Ratio) and $SHR@k$ (the sampling version), it is beneficial to take a functional view of them. Let $\mathcal{R}$ be the random variable for the user's item rank, with probability mass function $Pr({\mathcal R}=R)$;  then, $HR@K$ is simply the empirical cumulative distribution of ${\mathcal R}$ ($\widehat{Pr}$): 
%Dong, change P_R to W_R
% \begin{small}
% \begin{equation}\label{eq:HR2}
% HR@K = \widehat{Pr}({\mathcal R} \leq K), \quad P_R = \widehat{Pr}({\mathcal R}=R)
% \end{equation}
% \end{small}
\begin{small}
\begin{equation}\label{eq:HR2}
HR@K = \widehat{Pr}({\mathcal R} \leq K), \quad W_R = \widehat{Pr}({\mathcal R}=R)
\end{equation}
\end{small}
For $SHR@k$, its direct meaning is more involved and will be examined below. For now, we note that $SHR@k$ is a function of $k$ varying from $1$ to $n$, where $n-1$ is the number of sampled items. 

Figure~\ref{fig:1a} displays the curves of functional fitting of empirical accumulative distribution $HR@K$ (aka the global top-$K$ Hit- Ratio, varying $K$ from $1$ to $N=3706$), for $6$ representative recommendation algorithms ($3$ classical and $3$ deep learning methods), on the MovieLens 1M dataset. To observe the performance of these methods more closely when the $K$ is small, we first highlight $K$ from $1$ to $350$, and then again from $60$ to $120$. 

Figure~\ref{fig:1b} displays the curves of functional fitting of function $SHR@K$ (the sampling top-$k$ Hit-Ratio, varying $k$ from $1$ to $n=100$) with $n-1$ samples, under sampling with replacement, for the same $6$ representative recommendation algorithms on the same dataset. 
Similarly, we highlight $k$ from $1$ to $10$, and then again from $2$ to $5$. 

How can the sampling Hit-Ratio curves help to reflect what happened in the global curves? Before we consider the more detailed relationship between them, we introduce the following results:
%Dong, change P_R to W_R, change r^R equation to p^R, fix missing HR_2
% \begin{theorem}[Sampling Theorem]
% \label{theorem1}
% Let us assume we have two global Hit-Ratio curves (empirical cumulative distribution), $HR_1@K$ and $HR_2@K$, and assume one curve dominates the other one, i.e., $HR_1@K \geq HR@K$ for any $1\leq K \leq N$; then, for their corresponding sampling curve at any $k$ for any size of sampling, we have 
% $$E(SHR_1@k) \geq E(SHR_2@k)$$ 
% \end{theorem}
\begin{theorem}[Sampling Theorem]
\label{theorem1}
Let us assume we have two global Hit-Ratio curves (empirical cumulative distribution), $HR_1@K$ and $HR_2@K$, and assume one curve dominates the other one, i.e., $HR_1@K \geq HR_2@K$ for any $1\leq K \leq N$; then, for their corresponding sampling curve at any $k$ for any size of sampling, we have 
$$E(SHR_1@k) \geq E(SHR_2@k)$$ 
\end{theorem}
\begin{proof}
\sloppy Recall Equation~\ref{eq: ESHR_def0}: $E(SHR@k)=\sum_{R=1}^N W_R \cdot p^R =\sum_{u=1}^M Pr(r^u \leq k)$.  Let us assign each user $u$ the weight $Pr(r^u \leq k)$ for both curves, $HR_1$ and $HR_2$. Now, let us build a bipartite graph by connecting any $u$ in the $HR_1$ with user $v$ in $HR_2$, if $R_u \leq R_v$. We can then apply Hall's marriage theorem to claim there is a one-to-one matching between users in $HR_1$ to users in $HR_2$, such that $R_u \leq R_v$, and $Pr(r^u \leq k) \geq Pr(r^v \leq k)$.  (To see that, use the fact that $\sum_{R=1}^K W^{(1)}_R \geq \sum_{R=1}^K W^{(2)}_R$, where $W^{(1)}_R$ and $W^{(2)}_R$ are the empirical probability mass distributions of user-ranks, or equivalently, $\sum_{R=K}^N W^{(1)}_R \leq \sum_{R=K}^N W^{(2)}_R$. Thus, any subset in $HR_1$ is always smaller than its neighbor set $N(HR_1)$ in $(HR_2$).  Given this, we can observe that the theorem holds.  
\end{proof}
% \begin{proof}
% \sloppy Recall $E(SHR@k)=\sum_{R=1}^N P_R \cdot Pr(r^R \leq k)=\sum_{u=1}^M Pr(r^u \leq k)$.  Let us assign each user $u$ the weight $Pr(r^u \leq k)$ for both curves, $HR_1$ and $HR_2$. Now, let us build a bipartite graph by connecting any $u$ in the $HR_1$ with user $v$ in $HR_2$, if $R_u \leq R_v$. We can then apply Hall's marriage theorem to claim there is a one-to-one matching between users in $HR_1$ to users in $HR_2$, such that $R_u \leq R_v$, and $Pr(r^u \leq k) \geq Pr(r^v \leq k)$.  (To see that, use the fact that $\sum_{R=1}^K P^{(1)}_R \geq \sum_{R=1}^K P^{(2)}_R$, where $P^{(1)}_R$ and $P^{(2)}_R$ are the empirical probability mass distributions of user-ranks, or equivalently, $\sum_{R=K}^N P^{(1)}_R \leq \sum_{R=K}^N P^{(2)}_R$. Thus, any subset in $HR_1$ is always smaller than its neighbor set $N(HR_1)$ in $(HR_2$).  Given this, we can observe that the theorem holds.  
% \end{proof}

The above theorem shows that, under the strict order of global Hit-Ratio curves (though it may be quite applicable for searching/evaluating better recommendation algorithms, such as in Figure~\ref{figure:ml-1m:correspondence}), sampling hit ratio curves can maintain such order. 

%Dong, fix wrong notation, SHR to HR@(K=k)
However, this theorem does not explain the stunning similarity, shapes and trends shared by the global and their corresponding sampling curves. Basically, the detailed performance differences among different recommendation algorithms seem to be well-preserved through sampling. 
However, unless $n\approx N$, $SHR@k$ does not correspond to $HR@(K=k)$ (as in what is being studied by Rendel~\cite{rendle2019evaluation}).
%Dong, confuse equation

Those observations hold on other datasets and recommendation algorithms as well, not only on this dataset. 
Thus, intuitively and through the above experiments, we may conjecture that it is the overall curve $HR@K$ that is being approximated by $SHR@k$. Since these functions are defined on different domain sizes $N\ vs\ n$, we need to define such approximation carefully and rigorously. 

% The core of this paper is to investigate the relationship between the global top $K$ hit ratio $HR@K$  and sampling top $k$, $SHR@k$. We first note that $HR@K$ is a basically the cumulative distribution function of $K$ vary from $1$ to $N$ (the total number of items), and $SHR@k$ is a function of $k$ vary from $1$ to $n$, where $n-1$ is the number of sampled items. 
\subsection{Mapping Function $f$}
To explain the similarity between the global and sampling top-$k$ Hit-Ratio curves, we hypothesize that there exists a function $f(k)$ such that the relation $SHR@k \approx HR@ f(k)$ holds for different ranking algorithms on the same dataset. In a way, the sampling metric $SHR@k$ is like ``signal sampling''~\cite{rao2018signals}, where the global metrics between top $1$ to $N$ are sampled (and approximated) at only $f(1)<f(2)<\cdots<f(n)$ locations, which corresponds to $SHR@k$ ($k=1,2, \cdots, n$). In general, $f(k)\neq k$ (when $n << N$) (~\cite{rendle2019evaluation}).

% We note that in , Rendle considers and rejects the case for $f(k)=k$ as being inconsistent. 

In order to identify such a mapping function, let us take a look at the error between $SHR@k$ and $HR@ f(k)$ : $|SHR@k - HR@f(k)|$ 
\begin{small}
\begin{equation}\label{eq:error}
    \begin{split}
       & \le |SHR@k - E[SHR@k]| + |E[SHR@k] - HR@f(k)|
    \end{split}
\end{equation}
\end{small}
Thanks to the Hoeffding's bound, we observe, 
\begin{small}
\begin{equation*}
    Pr(|SHR@k - E[SHR@k]|\ge t) \le 2\exp(-2Mt^2)
\end{equation*}
\end{small}
This can be a rather tight bound, due to the large number of users in the population.
For example, if $M = 30K, t = 0.01$:
\begin{small}
\begin{equation*}
    Pr(|SHR@k - E[SHR@k] \ge 0.01|\le 0.005
\end{equation*}
\end{small}
If we want to look at more closely, we may use the law of large numbers and utilize the variance in equation ~\ref{eq: ESHR_var} for deducing the difference between $SHR@k$ and its expectation. 
Overall, for a large user population, the sampling top-$k$ Hit-Ratio will be tightly centered around its mean. Furthermore, if the user number is, indeed, small, an average of multiple sampling results can reduce the variance and error. In the publicly available datasets, we found that one set of samples is typically very close to the average of multiple runs. 

%Dong, change\em to \bf
Given this, our problem is {\bf how to find the mapping function $f$, such as $|E[SHR@k] - HR@f(k)|$ can be minimized (ideally close to or equal to $0$.} Note that $f$ should work for all the $k$ (from $1$ to $n$), and it should be independent of algorithms on the same dataset. 

%In the next section, we will introduce a few options of $f$, and then we will experimentally study and compare these $f$ functions. 

% Given this, 

% should be $0$ for appropriate mapping. To establish $SHR@k \approx HR@ f(k)$,  it's equivalent to force the error to approximate to $0$. We force the second term to approach $0$ in the remaing part of this section and the first term will be discussed in the next section.

%\vspace*{-1.0ex}
\section{ Approximating Mapping Function $f$} 
\label{estimator}

%Dong, change P_u to p_u, basically, all use p_u; fix indicator equation
\noindent{\bf Baseline:}
To start, we may consider the following naive mapping function.
We notice that for any $n$, 
\begin{small}
$$E(X^u)=(n-1) p_u= (n-1) \frac{R_u - 1}{N - 1}=E(r_u)-1$$ 
\end{small}
When the sample $n$ is large, we simply use the indicator function ${\bf 1}_{E(r_u) \leq k}$ to approximate and replace $Pr(r_u \leq k)$. Thus,
\begin{small}
\begin{equation}\label{eq:fk1}
   R_u \leq \frac{k - 1}{n - 1}*(N-1) + 1 = f(k) 
\end{equation}
\end{small}

Since the indicator function (${\bf 1}_{E(r_u) \leq k}$) is a rather crude estimation of the CDF of $r_u$ at $k$, this only serves as a baseline for our approximation of the mapping function $f$. 

\begin{figure}[h]
    \centering
        \includegraphics[width=\linewidth, scale = 0.7]{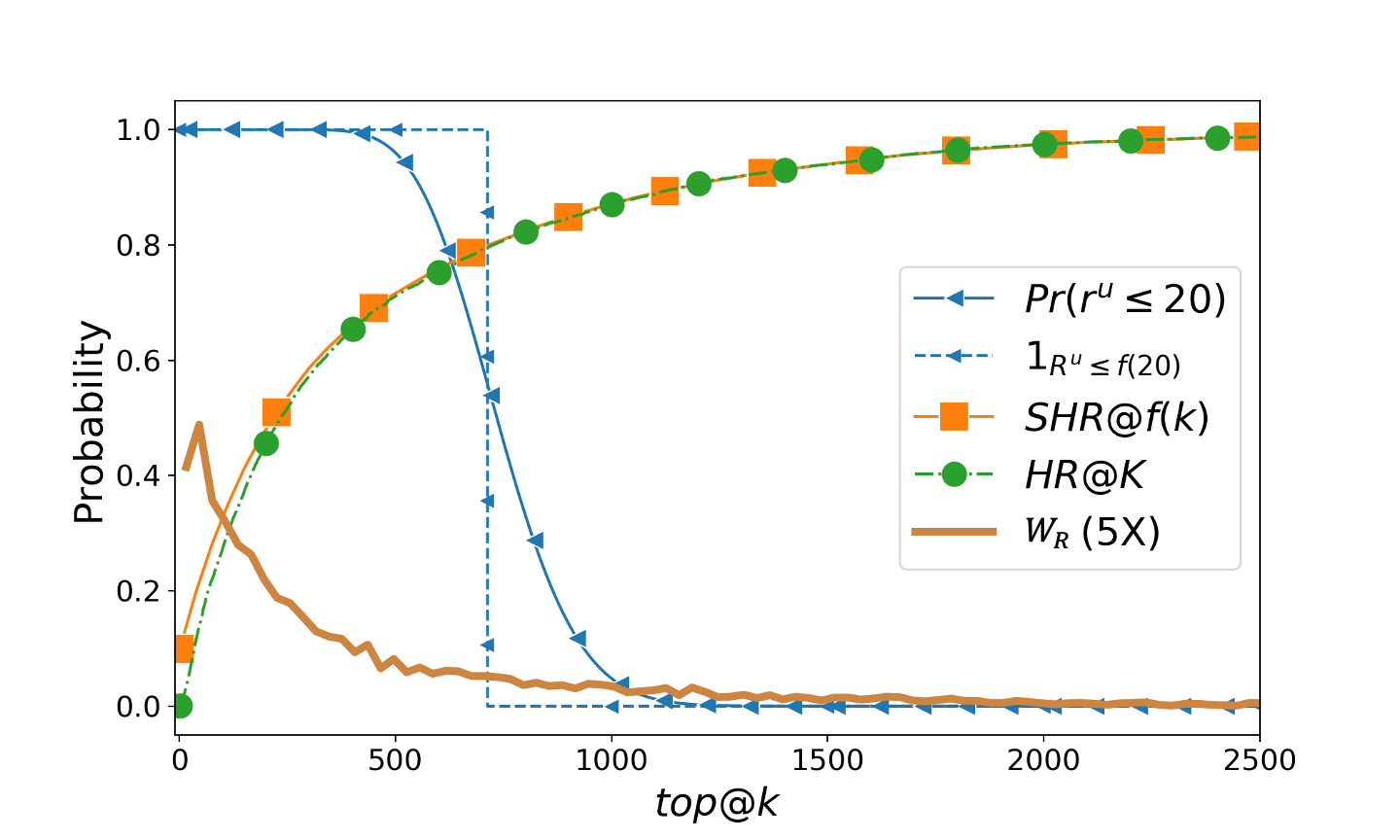}
    \caption{Curve Relationship. $HR@K$ is the top-K global hit-ratio;  $SHR@f(k)$ is the sampling top-k hit-ratio shown in global scale; $W_R(5X)$ is the empirical user ranking distribution, where we multiply by $5$ for displaying purpose.}
    \label{fig:unweighted_scalability}
\end{figure}

%Change P_R to W_R, P_u to p_u
\noindent{\bf Approximation Requirements:}
Before we introduce more carefully designed approximations of the mapping function $f$, let us take a close look of the expectation of the sampling top-$k$ Hit-Ratio $E(SHR@k)$ and $HR@f(k)$. Figure~\ref{fig:unweighted_scalability} shows how the user probability mass function $W_R$ works with the step function (indicator function) ${\bf 1}_{R_u\leq f(k)}$, and $p_u$ (assuming a hypergeometric distribution), to generate the global top-$K$ and sampling Hit-Ratios. 

%TODO, add more detailed caption

We make the following observations (as well as requirements):

\noindent{\em Existence of mapping function $f$ for each individual $HR$ curve:}
Given any $k$, assuming $HR@f(k)$ is a continuous cumulative distribution function (i.e., assuming that there is no jump/discontinuity on the CDF, and that $f(k)$ is a real value), then, there is $f(k)$ such that $HR@f(k)=E(SHR@k)$. 

In our problem setting, where $f(k)$ is integer-valued and ranges between $0$ and $N$, the best $f(k)$, theoretically, is 
\begin{small}
$$f(k)=\arg \min |HR@f(k)-E(SHR@k)|$$ 
\end{small}
\noindent{\em Mapping function $f$ for different $HR$ curves:}
Since our main purpose is for $SHR@k$ to be comparable across different recommendation algorithms, we prefer $f(k)$ to be the same for different $HR$ curves (on the same dataset).
Thus, by comparing different $SHR@k$, we can infer  their corresponding Hit-Ratio $HR$ at the same $f(k)$ location.  
Recall Figures~\ref{figure:ml-1m:correspondence} shows that the sampling Hit-Ratio curves are comparable with respect to their respective counterparts, and suggests that such a mapping function, indeed, may exist. 

%Dong, change P_R to W_R
But how does this requirement coexist with the first requirement of the minimal error of individual curves? We note that, for most of the recommendation algorithms, their overall Hit-Ratio curve $HR@K$, and the empirical probability mass function $W_R$ (Formula~\ref{eq: PR_def}), are actually fairly similar. 
From another viewpoint, if we allow individual curves to have different optimal $f(k)$, the difference (or shift) between them is rather small and does not affect the performance comparison between them, using the sampling curves $SHR@k$.
We will make this case more rigorously in Section~\ref{specific}, and we refer to this problem as the {\em sampling correspondence alignment problem}. 

In this section, we will focus on studying dataset-independent mapping functions, and we will discuss the algorithm-specific and dataset-specific mapping function in the next section.

%\vspace*{-1.0ex}
\subsection{Boundary Condition Approximation}
Consider that sampling with replacement, for any individual user, $X^u$ from equation \ref{eq:binom_dist}, obeys binomial distribution.  Apply the general case of bounded variables Hoeffding's inequality: 
\begin{small}
$$Pr(|X^u - E[X^u]|\ge t) \le 2e^{-\frac{2t^2}{n-1}} $$
\end{small}
%\begin{equation*}\label{eq:hoff0}
%\end{equation*}
since $r^u = X^u + 1$, and $E[r^u] = E[X^u]+1 = (n-1)p_u + 1$  :

% \begin{equation*}\label{eq:ru0}
%         Pr(|{r}^u - E[{r}^u]|\ge t)\le 2e^{-\frac{2t^2}{n-1}}
% \end{equation*}

\begin{small}
 \begin{equation}\label{eq:ineq_lr0}
    \begin{cases}
    &Pr(r^u \ge (n-1)p_u + 1 + t)\le 2e^{-\frac{2t^2}{n-1}} \\
    &Pr(r^u \le(n-1)p_u +1 - t)\le 2e^{-\frac{2t^2}{n-1}}
    \end{cases}
\end{equation}
\end{small}
The above inequalities indicate that $r^u$ is restricted around its expectation within the range defined by $t$. 

% =\sum\limits_{R = 1}^{N}P_R p^R - \delta(R\le f(k))]\\
%         &=\sum\limits_{R = 1}^{f(k)}P_R[Pr(r^R \le k) - 1]+ \sum\limits_{R = f(k)+ 1}^{N}P_R Pr(r^R \le k)\\
%         &
%Dong, change  P_R to W_R
The second term of error in equation \ref{eq:error} can be written as:
\begin{small}
\begin{equation}\label{eq:bound1}
    \begin{split}
        &E[SHR@k] - HR@f(k)\\
        &= - \sum\limits_{R = 1}^{f(k)}W_R\cdot Pr(r^R \ge k + 1)+ \sum\limits_{R = f(k)+ 1}^{N}W_R\cdot Pr(r^R \le k)
    \end{split}
\end{equation}
\end{small}
where, $r^R=r^u$ for $R_u=R$. 
For some relatively large $t$ (compared to $\sqrt{n-1}$), the probability in equation \ref{eq:ineq_lr0} can come extremely close to $0$. Based on this fact, if we would like to limit the first term $Pr(r^R\le k+1)$ to approach $0$, $k+1$ must be greater than $(n-1)p_u + 1+t$. And similar to the second term, we have:
% \begin{small}
%  \begin{equation*}\label{eq:bound2}
% %    \begin{cases}
%     &r^u \ge k+1 \ge (n-1) p_u + 1 + t) \quad
%   r^u \le k \le(n-1) p_u +1 - t)
% %$    \end{cases}
% \end{equation*}
% \end{small}

% Thus, we have: 
\begin{small}
 \begin{equation*}\label{eq:bound_lr2}
    \begin{cases}
    r^u \geq k+1 \ge (n-1)\frac{R-1}{N-1} + 1 + t), & R = 1,\dots,f^l(k) \\
   r^u \leq k \le(n-1)\frac{R-1}{N-1} +1 - t, & R =f^u(k) +1,\dots,N
    \end{cases}
\end{equation*}
\end{small}
where $f^l(k)$ and $f^u(k)$ are the lower bound and upper bound for $f(k)$, respectively. Explicitly, 
\begin{small}
 \begin{equation}\label{eq:bound_lr2}
    f^l(k) \le (k-t)\cdot\frac{N-1}{n-1} +1, \quad 
   f^u(k)\ge (k+t-1)\cdot\frac{N-1}{n-1}
\end{equation}
\end{small}
Given this, let the average of these two for $f$:
\begin{small}
\begin{equation}\label{eq:bound_mid}
     \boxed{f(k)= \lfloor\frac{f^l(k)+f^u(k)}{2} \rfloor=\lfloor (k-\frac{1}{2})\frac{N-1}{n-1}+\frac{1}{2} \rfloor}
\end{equation} 
\end{small}
Note that, although this formula appears similar to our baseline Formula~\ref{eq:fk1}, the difference between them is actually pretty big ($\approx \frac{1}{2}\frac{N-1}{n-1}$).  
As we will show in the experimental results, this formula is remarkably effective in reducing the error $|HR@f(k)-SHR@k|$. 
%\vspace*{-1.0ex}
%Dong, change P_R to W_R 
\subsection{Beta Distribution Approximation}
\begin{figure}[h]
    \centering
    \includegraphics[ width=\linewidth]{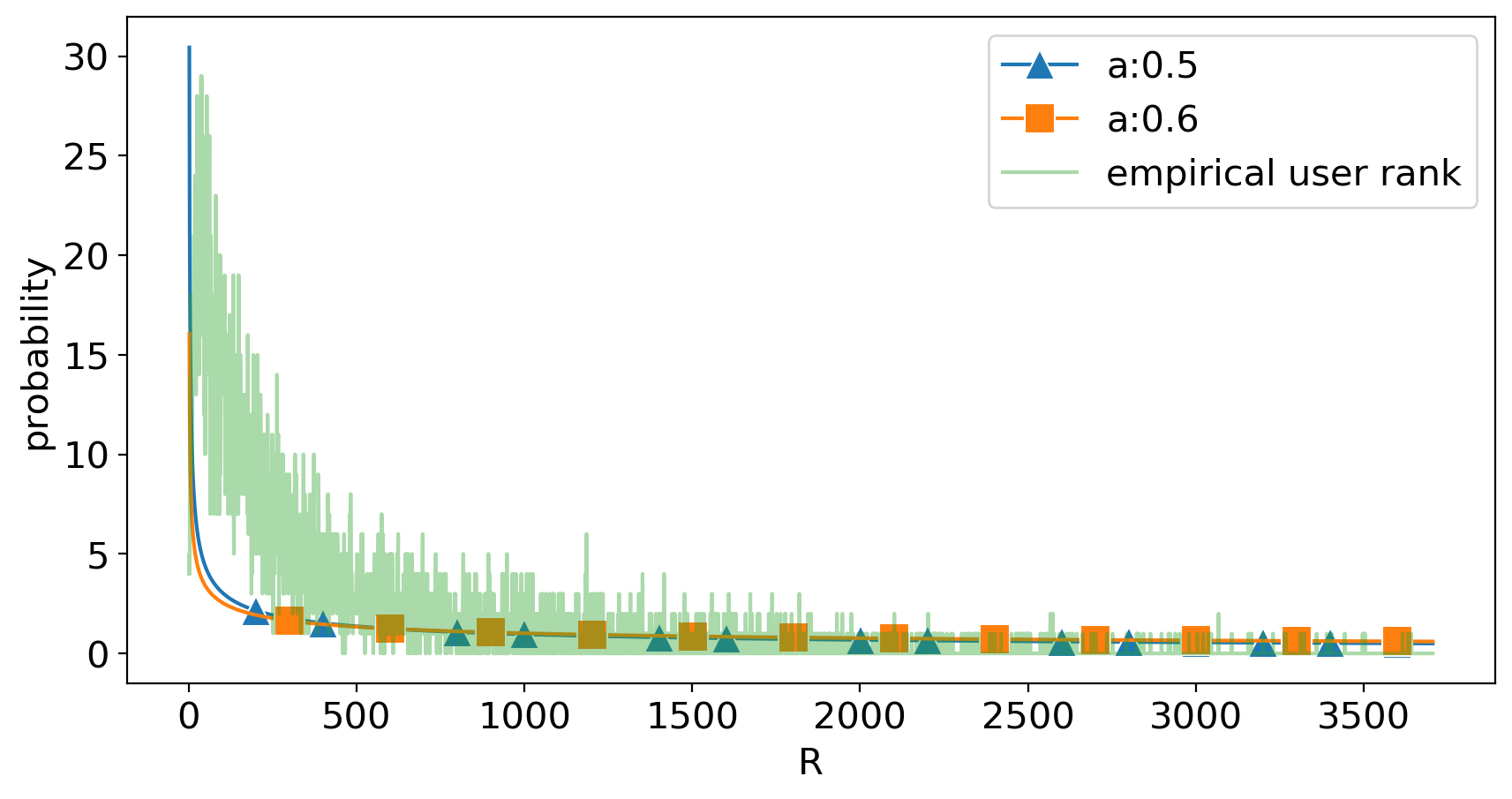}
    \caption{Beta distributions and empirical user rank distribution $W_R$}
    \label{fig:beta_user}
\end{figure}

In this approach, we try to directly minimize $HR@f(k) - E[SHR@k]$ to $0$, and this is equivalent to: 
\begin{small}
\begin{equation}\label{eq:lr0}
\begin{split}
&\sum\limits_{R = 1}^{N}{W_R \cdot {\bf 1}_{R\le f(k)}}= \sum\limits_{R = 1}^{N}{W_R\cdot Pr(r^R\le k)}
\end{split}
\end{equation}
\end{small}
\sloppy In order to get a closed-form solution of $f(k)$ from the above equation, we leverage the Beta distribution $Beta(a,1)$ to represent the user ranking distribution $W_R$, inspired by ~\cite{LiMG@Linguistic}: 
$W_R = \frac{1}{\mathcal{B}(a, 1)}(\frac{R -1}{N-1})^ {a-1} \frac{1}{N-1}$,  
where $a$ is a constant parameter and $\frac{1}{N-1}$ is the constant for discretized Beta distribution. Note that $\frac{R -1}{N-1}$ normalizes the user rank $R_u$ from $[1, N]$, to $[0, 1]$. Especially, when $a<1$, this distribution can represent exponential distribution, which can help provide fit for the $HR$ distribution. Figure~\ref{fig:beta_user} illustrates the Beta distribution fitting of $W_R$.
%Dong, change P_R to W_R 
%TODO, more detailed caption

% \begin{equation*}
%     p(x|a,b) = \frac{1}{\mathcal{B}(a,b)}x^{a-1}(1-x)^{b-1}
% \end{equation*}

% and make the assumption:
% \begin{equation}\label{eq: CR}
% p_R = \frac{1}{\mathcal{B}(a, 1)}(\frac{R -1}{N-1})^ {a-1}
% \end{equation}
% where $a$ is a constant parameter. The true distribution and the simulation are both plotted in figure[TBD].

%let $x = \frac{R - 1}{N - 1}$ then $\Delta x = \frac{1}{N-1}$.

%Dong, change P_R to W_R 
The left term of the equation \ref{eq:lr0} is denoted as $\mathcal{L}_k$
\begin{small}
\begin{equation*}\label{eq:left0}
\begin{split}
    &=  \sum\limits_{R = 1}^{N}{W_R \cdot {\bf 1}_{R\le f(k)}}  = \sum\limits_{R = 1}^{f(k)}{W_R}
    =\frac{1}{\mathcal{B}(a,1)}\sum\limits_{R = 1}^{f(k)}{(\frac{R -1}{N-1})^ {a-1}\cdot \frac{1}{N-1}} \\
    & = \frac{1}{\mathcal{B}(a,1)} \sum\limits_{x = 0}^{\frac{f(k) -1}{N -1} } {x^ {a-1}\cdot \Delta x} \quad \text{where}, x = \frac{R - 1}{N - 1}, \ and \Delta x = \frac{1}{N-1}, \\
    &\approx \frac{1}{\mathcal{B}(a,1)}\int_{0}^{\frac{f(k) -1}{N -1} }x^{a-1}dx 
     = \frac{1}{a\mathcal{B}(a,1)}[\frac{f(k) -1}{N - 1}]^a
\end{split}
\end{equation*}
\end{small}

% \sum\limits_{R = 1}^N{P_R Pr(r^R\le k)} 
%  = \sum\limits_{R = 1}^N{P_R\sum\limits_{i = 0}^{k-1}{\binom{n-1}{i}(\frac{R-1}{N-1})^i (1-\frac{R-1}{N-1})^{n-i-1}} }  \\
% & =

Considering sampling with replacement, then the right term is denoted as: 
\begin{small}
\begin{equation*} \label{eq:right0}
\begin{split}
&\mathcal{R}_k =  \sum\limits_{i = 0}^{k-1}{\binom{n-1}{i}\sum\limits_{R = 1}^N{W_R(\frac{R-1}{N-1})^i (1-\frac{R-1}{N-1})^{n-i-1}} }
\end{split}
\end{equation*}
\end{small}

Calculate the difference:
\begin{small}
\begin{equation*}\label{eq:right1}
\begin{split}
&\mathcal{R}_{k+1} -\mathcal{R}_k =  {\binom{n-1}{k}\sum\limits_{R = 1}^N{W_R(\frac{R-1}{N-1})^k (1-\frac{R-1}{N-1})^{n-1-k}} }\\
% &=  {\binom{n-1}{k}\sum\limits_{R = 1}^N{ \frac{1}{\mathcal{B}(a, 1)}(\frac{R-1}{N-1})^{a-1} (\frac{R-1}{N-1})^k (1-\frac{R-1}{N-1})^{n-1-k}} }\\
% &= \binom{n-1}{k}\frac{1}{\mathcal{B}(a, 1)}\sum\limits_{x = 0}^{1}{ x^{a+k-1} (1-x)^{n-1-k} \frac{1}{N-1}}\\
&\approx \binom{n-1}{k}\frac{1}{\mathcal{B}(a, 1)}\int_{x = 0}^{1} { x^{a+k-1} (1-x)^{n-1-k} dx} \\ 
& = \binom{n-1}{k}\frac{1}{\mathcal{B}(a, 1)}\mathcal{B}(a+k, n-k)
 =\frac{1}{\mathcal{B}(a, 1)} \frac{\Gamma(n)}{\Gamma(n+a)}\frac{\Gamma(k+a)}{\Gamma(k+1)}
\end{split}
\end{equation*}
\end{small}

Based on above equations: $\mathcal{L}_{k+1}-\mathcal{L}_{k}\approx \mathcal{R}_{k+1}-\mathcal{R}_{k}$, we have 
(we denote the mapping function as $f(k;a)$ for parameter $a$).

\begin{small}
\begin{equation}\label{eq:l2r}
 \begin{split}
        &[f(k+1;a)-1]^a-[f(k;a)-1]^a \\
        &=a[N-1]^a\binom{n-1}{k}\mathcal{B}(a+k, n-k)
        % &= a[N-1]^a \frac{\Gamma(n)}{\Gamma(n+a)}\frac{\Gamma(k+a)}{\Gamma(k+1)}
    \end{split}
\end{equation}
\end{small}

Then we have the following recurrent formula:
\begin{small}

\begin{equation}\label{eq:fk}
\boxed{
\begin{split}
    f(k+1;a)=&\Big[ a [N-1]^a \binom{n-1}{k}\mathcal{B}(a+k, n-k)\\
    &+ [f(k;a)-1]^a\Big]^{1/a} + 1
    \end{split}
}
\end{equation}

\end{small}
where we have $f(1)$ by considering $\mathcal{L}_1 = \mathcal{R}_1$:
\begin{small}
\begin{equation}\label{eq:f1}
f(1;a) = (N-1)[a\mathcal{B}(a,n)]^{1/a} + 1 
\end{equation}
\end{small}
%%\vspace*{-1.0ex}
\subsection{\bf Properties of Recurrent function $f$}
In the following, we enumerate a list of interesting properties of this recurrent formula of $f$ based on Beta distribution. 

\begin{lemma}[Location of Last Point]
For any $a$, all $f(n)$ converge to $N$: $f(n)=N$. 
\end{lemma}
\begin{proof}
We note that 
\begin{small}
\begin{equation*}
\begin{split}
    &\sum\limits_{k=0}^{n-1}{\binom{n-1}{k}\mathcal{B}(a+k, n-k)} 
    = \int_{0}^{1}{\sum\limits_{k=0}^{n-1}{\binom{n-1}{k}} t^{a+k-1}(1-t)^{n-k-1}dt} \\
    &= \int_{0}^{1}{ t^ {a-1}\left[\sum\limits_{k=0}^{n-1}{\binom{n-1}{k}} t^{k}(1-t)^{n-k-1} \right]dt} 
    = \int_{0}^{1}{ t^ {a-1}} = \frac{1}{a}
    \end{split}
\end{equation*}
\end{small}
Add up equation \ref{eq:l2r} from $k=1$ to $n-1$, we have:
\begin{small}
\begin{equation*}\label{eq:nN}
\begin{split}
&\frac{[f(n)-1]^a-[f(1)-1]^a}{a[N-1]^a} = \sum\limits_{k=1}^{n-1}{\binom{n-1}{k}\mathcal{B}(a+k, n-k)}\\
&=\frac{1}{a}-\binom{n-1}{0}\mathcal{B}(a,n) \quad 
% &\frac{[f(n)-1]^a-[N-1]^a \codt a\mathcal{B}(a,n)}{a[N-1]^a} = \frac{1}{a} - \mathcal{B}(a,n)\\
\text{then, we have } f(n) = N 
\end{split} 
\end{equation*}
\end{small}
\end{proof}
%\eproof

\noindent{\bf Uniform Distribution and Linear Map:} 
When the parameter $a=1$, the Beta distribution degenerates to the uniform distribution. 
From equations \ref{eq:f1} and \ref{eq:fk}, we have another simple linear map:
\begin{equation}\label{eq:uni_fk}
    f(k;a=1) = k\frac{N-1}{n} + 1
\end{equation}

Even though the user rank distribution is quite different from the uniform distribution, we found that this formula provides a reasonable approximation for the mapping function, and generally, better than the Naive formula \ref{eq:fk1}. 
More interestingly, we found that when $a$ ranges from $0$ to $1$ (as they express an exponential-like distribution), they actually are quite close to this linear formula.

\begin{figure}[h]
    \centering
    \includegraphics[width =\linewidth ]{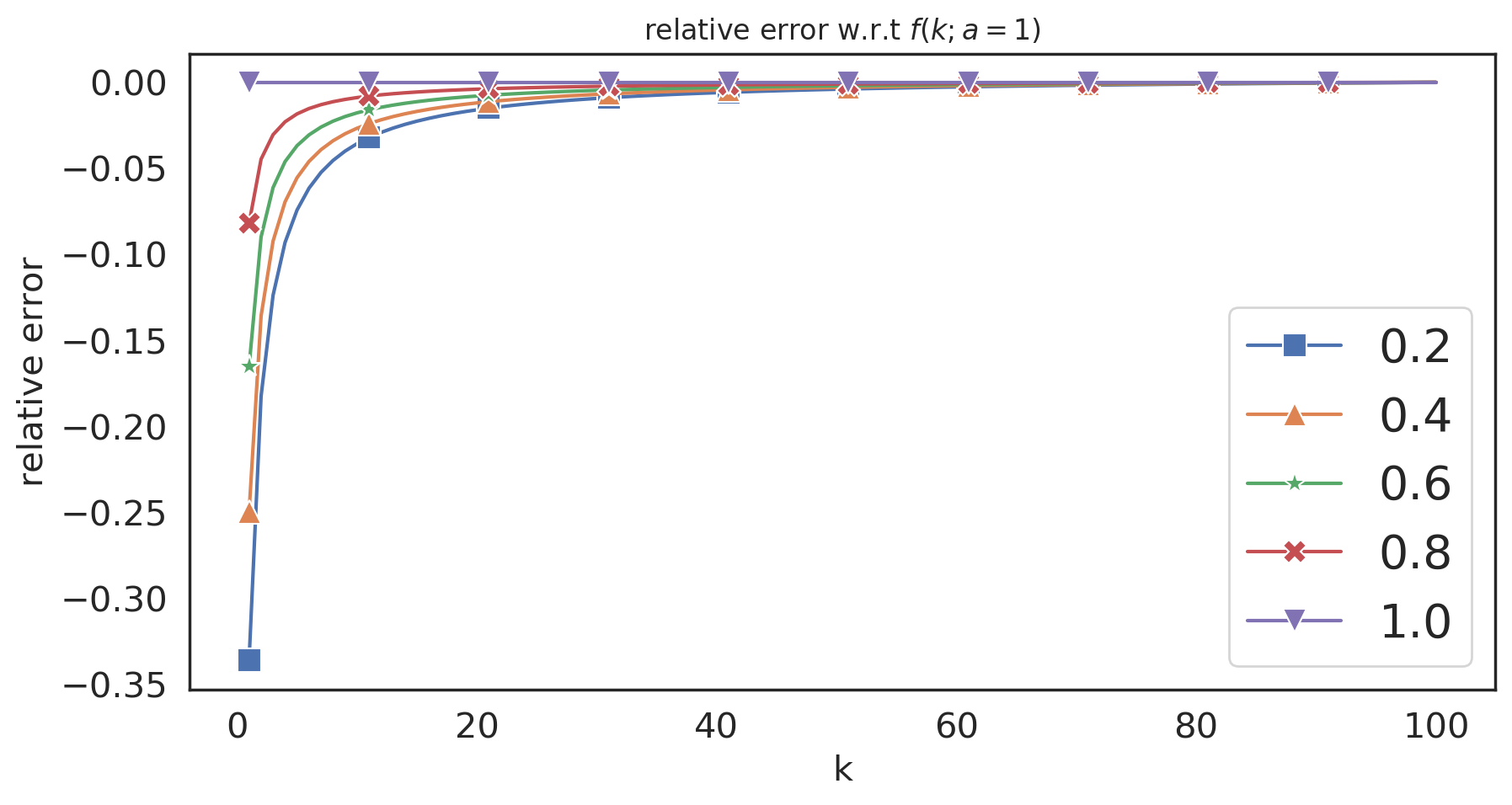}
    \caption{Relative error w.r.t. $f(k;a=1)$.ml-1m dataset, n= 100.}
    \label{fig:rela_a1}
\end{figure}

\noindent{\bf Approximately Linear:}
When we take a close look at the $f(k;a)$ sequences ($f(1;a), f(2;a), \cdots f(k;a)$ for different parameters $a$ from $0$ to $1$, \footnote{This also holds when $a>1$, but since the Hit-Ratio, aka the user rank distribution, is typically very different from these settings, we do not discuss them here.} we find that when $k$ is large, $f(k;a)$ all gets very close to $f(k;1)$ (the linear map function for the uniform distribution). Figures~\ref{fig:rela_a1} show the relative difference of all $f(k;a)$ sequences for $a=0.2,0.6,0.8$ with respect to $a=1$, i.e., $[f(k;a)-f(k;a=1)]/f(k;a=1)$. Basically, they all converge quickly to $f(k;a-1)$ as $k$ increases.

To observe this, 
  let us take a look at their $f(k)$ locations when $k$ is getting large. To simplify our discussion, let $g(k) = f(k)-1$, and then we have 
\begin{small}
\begin{equation*}\label{eq:LinearAp}
\begin{split}
     &[g(k+1)]^a - [g(k)]^a = a(N-1)^a\frac{\Gamma(n)}{\Gamma(n+a)}\frac{\Gamma(k+a)}{\Gamma(k+1)}\\
     &[\frac{g(k+1)}{g(k)}]^a = 1 +[\frac{(N-1)k}{g(k)n}]^a\frac{a}{k} \\
     & \text{when $n$ and $k$ are large,}  \lim_{n\rightarrow \infty}{\frac{\Gamma(n)}{\Gamma(n+a)}} = \frac{1}{n^a}  \\
     &[\frac{g(k+1)}{g(k)}] = \left(1 +[\frac{(N-1)k}{g(k)n}]^a\frac{a}{k}\right)^{1/a}\approx 1 +[\frac{(N-1)k}{g(k)n}]^a\frac{1}{k}
\end{split}
\end{equation*}
\end{small}
When $g(k) = (N-1)\frac{k}{n}$, the above equation holds $\frac{g(k+1)}{g(k)} = 1+\frac{1}{k}$, and this suggests they are all quite similar to the linear map $f(k;a=1)$ for the uniform distribution. 

By looking at the difference $f(k;a)-f(k-1;a)$, we notice we will get very close to the constant $\frac{N-1}{n}=f(k;1)-f(k-1;1)$ even when $k$ is small. To verify this, let $y_{k+1}$
\begin{small}
$$ = [f(k+1;a)-1]^a - [f(k;a) - 1]^a = a[N-1]^a\frac{\Gamma(n)}{\Gamma(n+a)}\frac{\Gamma(k+a)}{\Gamma(k+1)}$$
\end{small}
Then we immediately observe: 
\begin{small}
\begin{equation*}
\begin{split}
    &\frac{y_{k+1}}{y_k}=\frac{
    a[N-1]^a\frac{\Gamma(n)}{\Gamma(n+a)}\frac{\Gamma(k+a)}{\Gamma(k+1)}}{a[N-1]^a\frac{\Gamma(n)}{\Gamma(n+a)}\frac{\Gamma(k-1+a)}{\Gamma(k)}}
    =1 + \frac{a-1}{k}
    \end{split}
\end{equation*}
\end{small} 
Thus, after only a few iterations for $f(k;a)$, we have found that their (powered) difference will get close to being a constant. 

%TODO, more detailed captions

% \subsection{Hypergeometric Distribution Discussion}

% \section{Large Population Effect}
% \subsection{Large Population Tight Expectation}
% The error in equation \ref{eq:error} can be upper bounded by right side. In this subsection, we  apply Hoeffding's Inequality to the first term.
% Revisit equation \ref{eq: ESHR0}, 

% \subsection{User Sampling}

% We sample $m$ users from the total user set and calculate the:
% \begin{equation}\label{eq:SSHR}
%     SSHR@k = \frac{1}{m}\sum\limits_{u=1}^{m}{Z_u}
% \end{equation}
% \begin{equation}
% \begin{split}
%     |SSHR@k - E[SHR@k]|\le |SSHR@k - SHR@k|\\
%     +|SHR@k - E[SHR@k]
%     \end{split}
% \end{equation}

% \begin{equation}
%     Pr(SSHR@k - SHR@k \ge \epsilon)\le \exp{(-2n\epsilon^2)}
% \end{equation}
% let $\epsilon = 0.02$, $n\ge 3700$ then we have $Pr \le 0.05 $

% \section{Re-Rank}

\section{Dataset-Specific Mapping Function}
\label{specific}
In the last section, we introduced some generic mapping functions. which can be used across different datasets. 
As we will show in Section~\ref{experiments}, choosing some generic $a$, e.g.,  $a=0.5$, can provide quite accurate approximation; the differences between $HR@f(k)$ and $SHR@k$ are quite small. However, why can such a generic $a$ be so effective? Furthermore, can we design a better mapping function which can leverage the inherent characteristics of recommendation algorithm performance on different datasets? 

By answer these questions, we found some further interesting properties of the sampling Hit-Ratio metrics: They can be a rather robust (or safe) measure to select the best algorithm! In layman's terms, if the sampling metric $SHR$ shows a good improvement of an algorithm over others, then there is a good chance that it may perform even better in the global top-$K$ Hit-Ratio metrics. 
Conversely, if an algorithm under-performed in the sampling measure, it may be even worse in the global measure. 

%Dong, change P_R to W_R
\subsection{Optimizing $a$ (Algorithm-Specific Mapping)}
Recall that we use $Beta(a,1)$ to fit $W_R$.  
Since $W_R$, the accumulative distribution of which is $HR$, is clearly unknown to us, we will try to use $SHR$ to represent it instead. But how does it relate to $HR$? Based on our earlier discussion, it can be approximated as $HR@f(k)\approx SHR@k$. 
Especially, if we leverage the $Beta$ distribution, when $a$ is given, we can use the aforementioned $f(k;a)$ for our purpose. Once we have such mapping, we can then use $SHR@k$ (more precisely  $\widehat{HR}@f(k)=SHR@k$ ), to help fit the beta distribution and consequently find a new parameter $a$.

Given this, we can utilize the following iterative procedure to identify the optimal $a$, which uses a maximal likelihood approach to help fit the beta distribution:
\begin{equation}\label{recursive}
    \begin{split}
        \boxed{a^{(i+1)} = \frac{-M}{\sum\limits_{u=1}^{M}{\ln(\frac{f(r^u;a^{(i)})-1}{N-1})}}}
    \end{split}
\end{equation}
where ${\mathcal L} := \prod_{u = 1}^{M}{a(\frac{f(r^u|a)-1}{N-1})^{a-1}\frac{1}{N-1}}$ is the likelihood function based on Beta distribution, and if we take the derivative of the log-likelihood, we have the above formula as the optimal parameter $a$.
We can start with any reasonable $a$, such as $a=1$ or $a=0.5$. 
% This is a continuous function of $a$ bounded between $0$ and $1$, and the fixed point theorems~\cite{fixpoint} can be used to verify its convergence property.  In addition,
This procedure can be considered as a simplified EM algorithm. Our experiments show it converges very quickly (in two or three iterations) to some fixed point.  
%\vspace*{-1.0ex}
\subsection{Sampling Correspondence Alignment Problem and its Remedy}

Now, for different recommendation algorithms on the same data with the same sampling size $n$, each produces their corresponding sampling Hit-Ratio $SHR@k$, and each will produce different parameters $a$ using the above method (Formula~\ref{recursive}), which leads to a different mapping function $f(k;a)$. This leads to the {\em sampling correspondence alignment problem}: {\em 
for any fixed $k$ under sampling, different algorithms' $SHR@k$ (with different $a$) measures their corresponding $HR$ at different location $f(k;a)$. }
Given this, can sampling $SHR@k$ still be meaningful for performance evaluation? 

\noindent{\bf Remedy $\# 1$: The difference between $a$ is very small:} Through extensive experimental evaluation, we found that on the same dataset, the optimal $a$ of different recommendation algorithms are actually very close to one another (See Table~\ref{tab:betaparameters}, sampling size $n=100$). In most of these datasets, their $a$'s difference is within $0.01$. 

\begin{scriptsize}
\begin{table}[h]
  \caption{Beta Parameter}
  \vspace{-4.0ex}
  \label{tab:betaparameters}
  \begin{tabular}{lcccc}
    \toprule
    \textbf{Model}
    &
    \textbf{ml-1m}&
    \textbf{yelp}&
    \textbf{pinterest-20}&
    \textbf{citeulike}\\
    \midrule
\textit{NeuMF}&0.3685 & 0.3059&0.2820 &0.2601\\
\textit{MultiVAE}&0.3681 &0.2977 & 0.2764&0.2449\\
\textit{EASE}&0.3684 & 0.2972&0.2806 &0.2532\\
\textit{itemKNN}& 0.4079&0.2885 &0.2782 &0.2519\\
    \bottomrule
  \end{tabular}
\end{table}
\end{scriptsize}

%How such small difference manifest in the mapping function $f(k;a)$? 

\begin{figure}[h]
    \centering
    \includegraphics[ width=\linewidth]{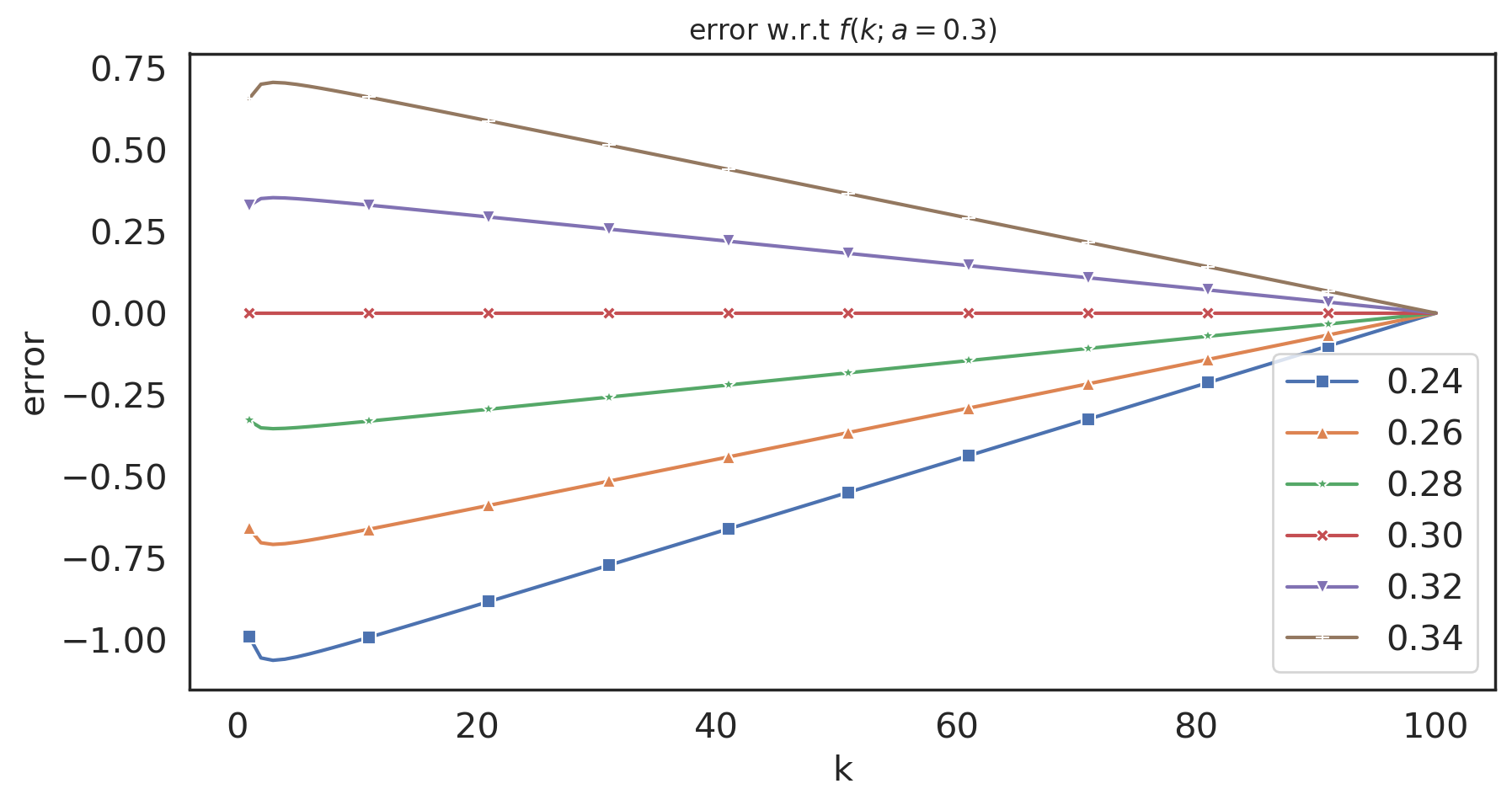}
    \caption{Error of $f(k;a)$ w.r.t $f(k;0.3)$, $a$ ranges from $0.24$ to $0.34$. ml-1m dataset, n = 100.}
    \label{fig:error3}
\end{figure}

\noindent{\bf Remedy $\# 2$: The difference between $f(k;a)$ for slightly different $a$ is very small:} When we put slightly different $a$ into the mapping function, and observe their difference, $f(k;a)-f(k;a^\prime)$ is also very small. Figure~\ref{fig:error3} shows the the difference of mapping functions  for $a$ ranges from $0.24$ to $0.34$ with the mapping function at $a=0.3$, i.e., $f(k;a)-f(k;a=0.3)$. We see that their absolute location difference is less than $1$; i.e., with the original sampling location on the scale from $1$ to $N$, for the same $k$, their correspondence location difference is less than $1$. Thus, we generally can use any of the $a$ obtained from one of the recommendation algorithms on a dataset as the choosing parameters for all the recommendation algorithms. In fact, this also suggests the parameter $a$ is an inherent parameter for each dataset when using the existing (competitive) recommendation algorithms. 
Thus, we can have dataset-specific $a$ (without worrying the algorithm-specific $a$). 

%TODO more detailed captions
\begin{figure}[h]
    \centering
    \includegraphics[ width=\linewidth]{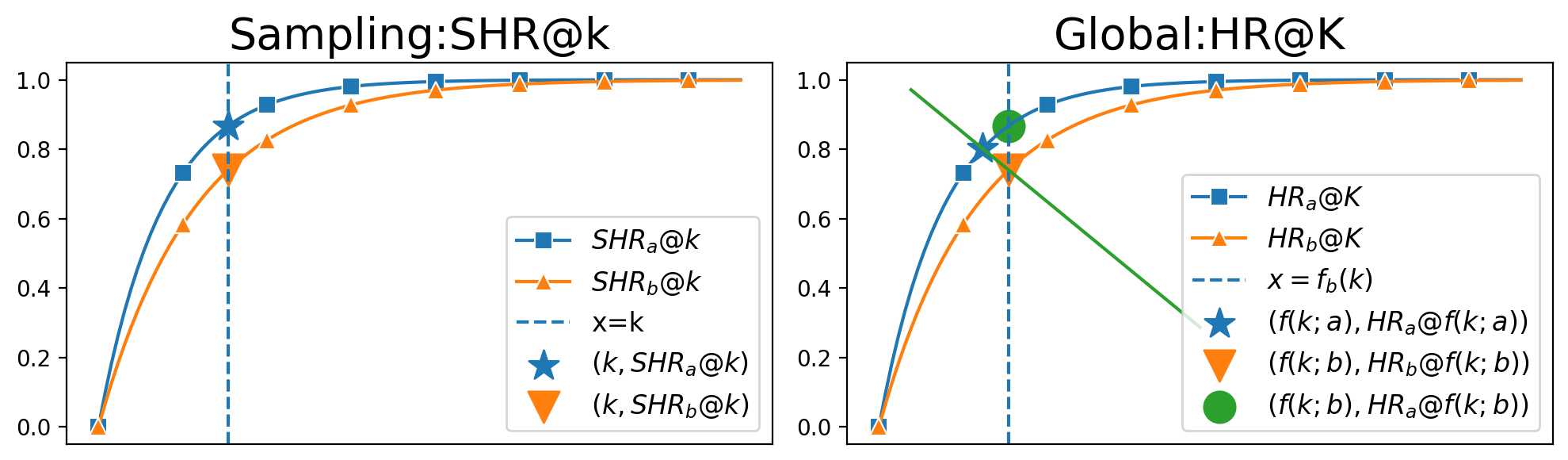}
    \caption{Sampling Effect}
    \vspace{-4.0ex}
    \label{fig:effect}
\end{figure}

\noindent{\bf Remedy $\# 3$: When $a$ is smaller, so is $f(k;a)$:} An interesting discovery can be made by observing Figures ~\ref{fig:error3} and ~\ref{fig:rela_a1}: When $a>b$, $f(k;a) \leq f(k;b)$ for any $k$. This holds even if $a$ and $b$ are not close.  When $a$ and $b$ are closer, then their difference has become smaller; but as long as $a$ is larger than $b$, the mapping function $f(k;a)$ always corresponds to an earlier location than $f(k;b)$. Intuitively, when $a$ is smaller, this corresponds to more users with higher rank, i.e., their Hit-Ratio $HR@K$ (accumulative distribution) under beta distribution is consistently better than the larger $b$. 
Mathematically, this is equivalent to saying that $f(k;a)$ is a monotone increasing function with respect to $a$ (for $a>0$). Even though we can numerically observe this, the rigorous proof of its monotonicity remains an open problem. 

%(due to the complexity of the its derivative). 

The implication of such a monotone property is quite interesting and likely useful: If the sampling metrics show a good improvement of an algorithm over another algorithm, then there is a good chance that it may perform even better in the global top-$K$ Hit-Ratio metric, as it may actually correspond to an earlier (or smaller) $f(k)$: assuming $SHR_a@k > SHR_b@k$, and $f(k;a)<f(k;b)$,   
\begin{small}
$$HR_a@f(k;b) \geq HR_a@f(k;a)\approx SHR_a@k \geq  SHR_b@k \approx HR_b@f(k;b) $$
\end{small}
Figure~\ref{fig:effect} illustrates this effect. 
This helps explain why the sampling Hit-Ratio is very effective in choosing the winner (or loser) of different recommendation algorithms. They ensure the correct prediction when two recommendation algorithms perform very differently (not  covered by Remedy $\# 2$). 

%The detailed evaluation is in Section~\ref{experiments}.

% Furthermore, no matter what mapping function $f$ we use, this holds: {\em for the same $k$, its correspondence location of the higher global Hit-Ratio tends to be smaller}.  

% (In this approach, we will consider to explicitly fit $Beta(a,1)$  
% As we described above in equation \ref{eq: CR}, the user rank satisfy the beta distribution:

% $$R \sim a(\frac{R-1}{N-1})^{a-1}$$

% For each user rank in sample $r_u$, the map function will give a corresponding $R^\prime_u = f(r_u|a)$

%\input{text/eval.tex}
\section{Experimental Results}
\label{experiments}

\begin{figure*}[t]
\begin{subfigure}{0.45\textwidth}
    \includegraphics[width=\textwidth, scale = 0.8]{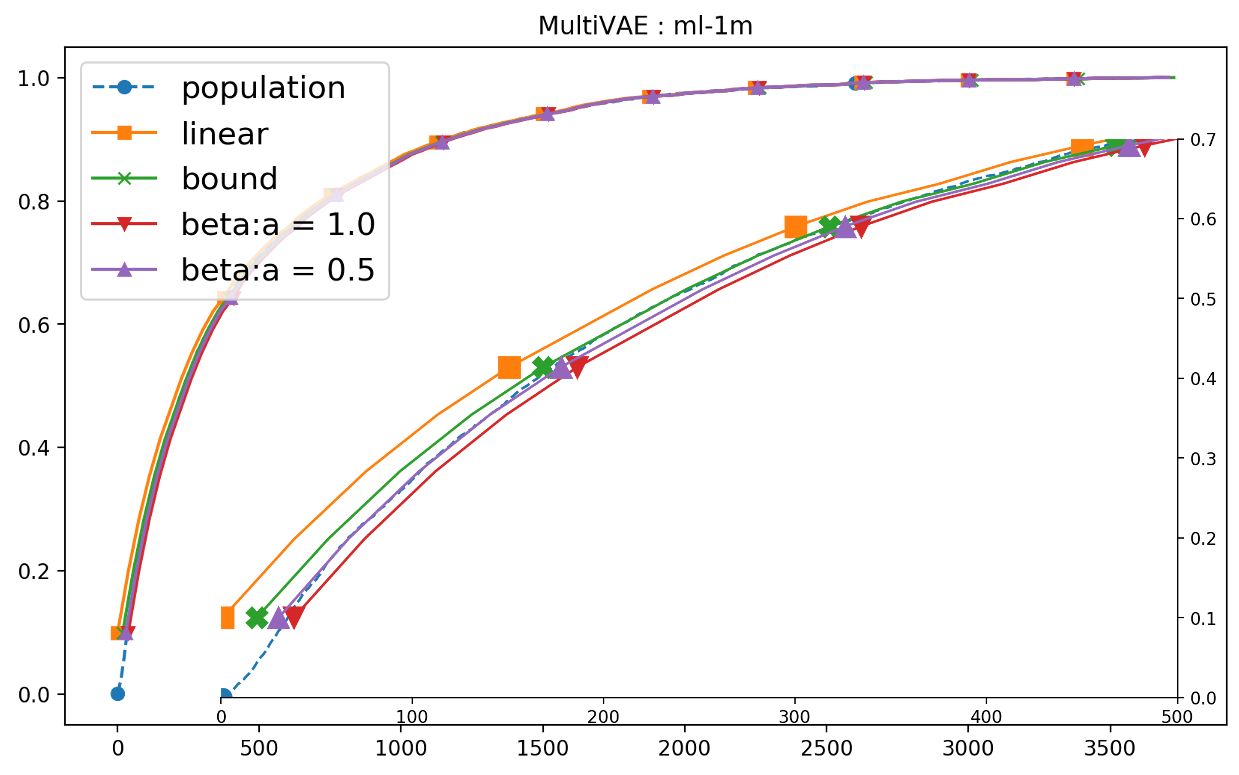}
    \caption{Hit Ratios from the Global Top-$K$. Except the "population" curve, all others map the sample hit-ratio to global scale by different mapping functions.}
    \label{figure:ml-1m:global}
\end{subfigure}
\hspace{0.05\textwidth}
\begin{subfigure}{0.45\textwidth}
    \includegraphics[width=\textwidth, scale = 0.8]{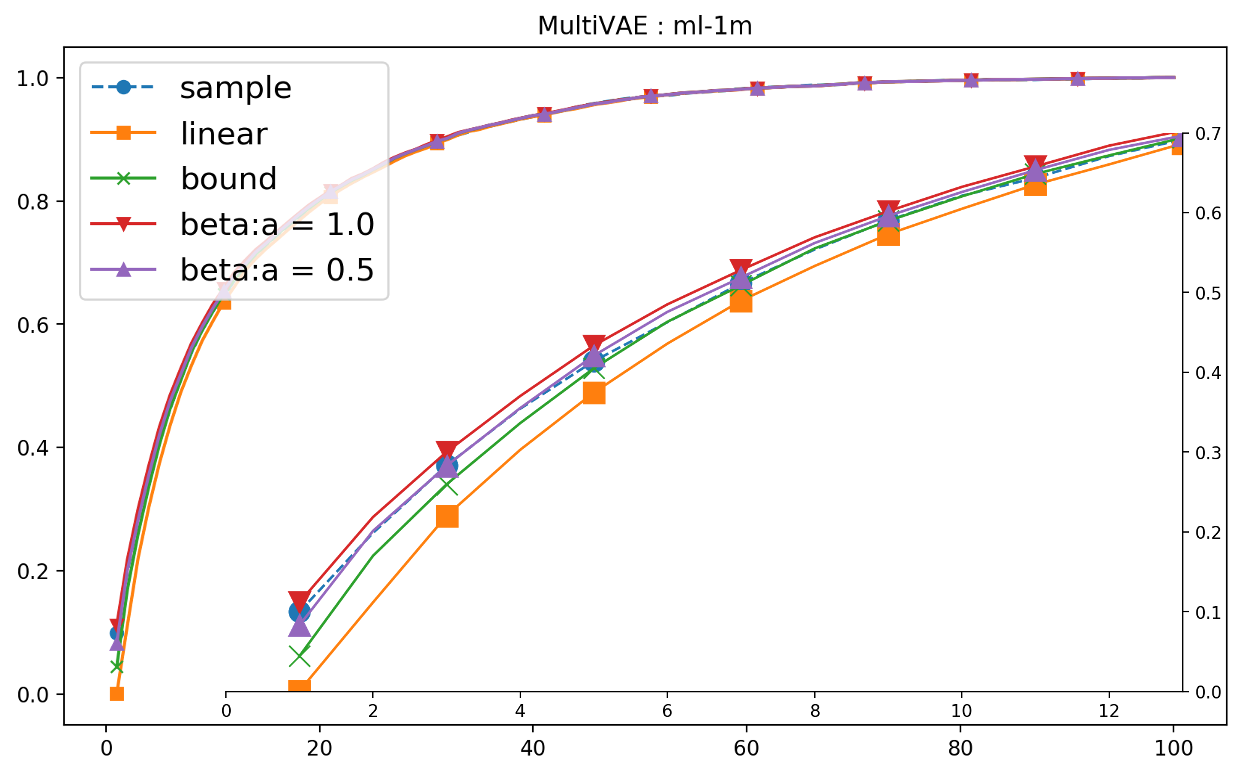}
    \caption{Hit Ratios from the Sampling Top-$k$. Except the "sample" curve, all others map the global/population hit-ratio to sample scale by different mapping functions.}
    \label{figure:ml-1m:sample}
\end{subfigure}
\caption{The right part of each figure is the zoom in version which display the relation of curves more clear.dataset setting: ml-1m, n = 100}
\vspace{-2.0ex}
\label{figure:ml-1m:Performance}
%\vspace*{-3.0ex}
\end{figure*}

In this section, we experimentally study the sampling hit ratio $SHR@k$ and its corresponding global hit ratio $HR@K$ through different mapping functions $f$. Specifically, we aim to answer: 

%the following questions:

\begin{itemize}
    \item (Question 1) How does the dataset-independent mapping function $f$ help align $SHR@k$ with respect to $HR@f(k)$? 
    \item (Question 2) How do different factors, such as the top-$k$ location, varying the effect of the sampling scheme, and the sampling size affect the results?
    \item (Question 3) How does the algorithm-specific mapping function $f$  compare with other mapping functions? 
    \item (Question 4) How can the random sampling hit ratios be used to identify the winners of recommendation algorithms with respect to the corresponding global hit ratio?
\end{itemize}

% $(which potentially provides the lower bound for a specific algorithm on a given dataset, based on Beta distribution)

In this section, we only focus on Question 1 and Question 4. The discussion of Question 2 and 3, together with their experimental results and the experimental setup will be put into the Appendix~\ref{ap:ex}. Due to the space limitation, we only report representative results here, and additional experimental results are openly available at \footnote{\url{https://github.com/dli12/KDD20-On-Sampling-Top-K-Recommendation-Evaluation}} .

%---------------- dataset--
% ml-1m \footnote{ \url{https://github.com/hexiangnan/neural_collaborative_filtering/tree/master/Data}} ~\cite{he2017neural}

% pinterest-20\footnote{\url{https://github.com/hexiangnan/neural_collaborative_filtering/tree/master/Data}} ~\cite{he2017neural}

% yelp\footnote{\url{https://github.com/duxy-me/ConvNCF/tree/master/Data}} ~\cite{he2018outer}

% citeulike\footnote{\url{https://github.com/tebesu/CollaborativeMemoryNetwork/tree/master/data}}
% ~\cite{ebesu2018collaborative}

% book-X\footnote{\url{http://www2.informatik.uni-freiburg.de/~cziegler/BX/}} ~\cite{Ziegler05}

%-----model---------%

\noindent{\bf Aligning Sampling and Global Hit Ratio $SHR$ and $HR$:} In this experiment, we provide two dual views of the alignment between sampling and global hit ratio (curves). Here, we report only the \texttt {MultiVAE} result on dataset \texttt{ml-1m} in Figures~\ref{figure:ml-1m:Performance}. We use four different dataset-independent mapping functions, the linear, bound, $beta@1$ and $beta@0.5$, for the curve alignment. Figure~\ref{figure:ml-1m:global} maps the sampling curve $SHR@k$ to the global top-$K$ view by mapping $SHR@k$ to location $f(k)$ in the population/global top $K$ view, and compares them with the global $HR@K$ (population curve). Figure~\ref{figure:ml-1m:sample} maps the global curve $HR@K$ to the sampling top-$k$ view by $HR@K$ to the location $k$ (where $K=f(k)$) in the sampling top $k$ view, and then compares them with the sample $SHR@k$ (sampling curve). We observe that both bound and $beta@0.5$ achieve the best results from both views. The same observation holds on other recommendation algorithms and datasets.

 % i.e., $\sum_{k=2}^10|HR@f(k)-SHR@k|/9$ and $\sum_{k=2}^10(|HR@f(k)-SHR@k|/SHR@k)/9$.
 
\noindent{\bf Predicting Winners (and Relative Performance):}
In Table~\ref{tab:consist}, we demonstrate the effectiveness of using sampling hit ratio $SHR@k$ to predict the recommendation algorithm performance when using the global hit ratio $HR@f(k)$.  We compare the performance of the three most competitive recommendation methods on the four commonly used datasets. We also vary the $k$ from $1$ to $50$ for sampling size $n-1=99$. Throughout all these cases, $SHR@k$ and $HR@f(k)$ all {\em consistently} predict the same and correct winners. Specifically, if an algorithm has the highest $SHR@k$, then it has the highest $HR@f(k)$ as well.  In fact, their relative orders are also mostly consistent besides the winners. 

\begin{scriptsize}

\begin{table*}[th]
    \caption{Predicting Winners (and Relative Performance)}
    \label{tab:consist}
    %\vspace*{-2.0ex}
    \begin{minipage}{0.48\textwidth}
      \caption*{ml-1m}
      \vspace*{-4.0ex}
      \centering
    \begin{tabular}{|l|ccc|ccc|ccc|}
    \hline
    &
    \multicolumn{3}{c|}{\textbf{NeuMF}}&
    \multicolumn{3}{c|}{\textbf{MultiVAE}}&
    \multicolumn{3}{c|}{\textbf{EASE}}\\
    \hline
    k&SHR&bound& $\mathcal{B}$@.5 &SHR&bound&$\mathcal{B}$@.5&SHR&bound&$\mathcal{B}$@.5\\
    \hline
1	&0.208&	0.150&	0.205&		0.211&	0.160&	0.216 &\textbf{0.223}&	\textbf{0.173}	&\textbf{0.232}	\\
2&	0.326&	0.311&	0.347	&	0.343	&0.319	&0.355 &\textbf{0.349}&	\textbf{0.334}&\textbf{	0.370}	\\
5&	0.548&	0.555&	0.566&		0.555&	0.560&	0.576&\textbf{ 0.564}&\textbf{	0.573}&	\textbf{0.587}	\\
10&	0.715&	0.726&	0.731&		0.717&	0.729&	0.733 &\textbf{0.720}	&\textbf{0.733	}&\textbf{0.737}\\
20&	0.850&	0.863&	0.864	&	\textbf{0.854}&	\textbf{0.866}&	\textbf{0.867}& 0.847&	0.859	&0.860	\\
50&	0.972&	0.979&	0.979	&	\textbf{0.972}&	\textbf{0.980}&\textbf{	0.979} &0.955&	0.961&	0.961	\\
    \hline
  \end{tabular}
    \end{minipage}\hfill
    \begin{minipage}{0.48\textwidth}
      \centering
        \caption*{yelp}
        \vspace*{-4.0ex}
 \begin{tabular}{|l|ccc|ccc|ccc|}
    \hline
    &
    \multicolumn{3}{c|}{\textbf{NeuMF}}&
    \multicolumn{3}{c|}{\textbf{MultiVAE}}&
    \multicolumn{3}{c|}{\textbf{EASE}}\\
    \hline
    k&SHR&bound&$\mathcal{B}$@.5&SHR&bound&$\mathcal{B}$@.5&SHR&bound&$\mathcal{B}$@.5 \\
    \hline
1&	0.234&	0.184&	0.248	&	0.262&	0.215&	0.283 &\textbf{0.275}&\textbf{0.228}&\textbf{0.295}\\
2	&0.369	&0.360	&0.392	&	0.404&	0.398&	0.429 &\textbf{0.418}&	\textbf{0.410}&	\textbf{0.442}	\\
5	&0.593	&0.600	&0.611	&	0.626&	0.635&	0.645 &\textbf{0.632}&	\textbf{0.642}&	\textbf{0.653}	\\
10&	0.753&	0.760&	0.764	&	0.775&	0.781&	0.783 &\textbf{0.777}&\textbf{	0.784}&	\textbf{0.787}	\\
20&	0.881&	0.885&	0.886	&	\textbf{0.892}&	\textbf{0.894}&	\textbf{0.895} &0.886&	0.888&	0.889\\
50	&\textbf{0.975}&\textbf{	0.976}&\textbf{	0.976}	&	0.974&	0.975&	0.975 &0.957&	0.957	&0.957\\
    \hline
  \end{tabular}
    \end{minipage}\hfill 
% \end{table*}
% \end{scriptsize}

% \begin{scriptsize}

% \begin{table*}[th]
%     \caption{Consistency}
    \begin{minipage}{0.48\textwidth}
      \caption*{pinterest-20}
      \vspace*{-4.0ex}
      \centering
    \begin{tabular}{|l|ccc|ccc|ccc|}
    \hline
    &
    \multicolumn{3}{c|}{\textbf{NeuMF}}&
    \multicolumn{3}{c|}{\textbf{MultiVAE}}&
    \multicolumn{3}{c|}{\textbf{EASE}}\\
    \hline
    k&SHR&bound& $\mathcal{B}$@.5 &SHR&bound&$\mathcal{B}$@.5&SHR&bound&$\mathcal{B}$@.5\\
   \hline
1&	0.273&	0.201&	0.278	&	\textbf{0.316}&	\textbf{0.241}	&\textbf{0.321} &0.289&	0.222&	0.298\\
2	&0.436&	0.409&	0.448	&\textbf{	0.479}&\textbf{0.456}&\textbf{0.495}& 0.452&	0.425&	0.465\\
5	&0.701&	0.708&	0.722&		\textbf{0.729}	&\textbf{0.735}	&\textbf{0.747}& 0.705&	0.712&	0.724\\
10&	0.874&	0.883&	0.886&		\textbf{0.887}	&\textbf{0.894}&	\textbf{0.897} &0.868&	0.877&	0.880\\
20&	0.965&	0.968&	0.968	&	\textbf{0.970}&	\textbf{0.973}	&\textbf{0.973}& 0.959&	0.963	&0.963\\
50	&0.993&	0.993&	0.993&		\textbf{0.994}&	\textbf{0.994}&	\textbf{0.994}& 0.989	&0.989&	0.989\\
    \hline
  \end{tabular}
    \end{minipage}\hfill
    \begin{minipage}{0.48\textwidth}
      \centering
        \caption*{citeulike}
        \vspace*{-4.0ex}
 \begin{tabular}{|l|ccc|ccc|ccc|}
    \hline
    &
    \multicolumn{3}{c|}{\textbf{NeuMF}}&
    \multicolumn{3}{c|}{\textbf{MultiVAE}}&
    \multicolumn{3}{c|}{\textbf{EASE}}\\
    \hline
    k&SHR&bound&$\mathcal{B}$@.5&SHR&bound&$\mathcal{B}$@.5&SHR&bound&$\mathcal{B}$@.5 \\
        \hline
1&	0.399&	0.427	&0.505	&	\textbf{0.488}&	\textbf{0.554}	&\textbf{0.615}& 0.464&	0.511	&0.577	\\

2&	0.583&	0.606&	0.632	&	\textbf{0.675}&	\textbf{0.707}	&\textbf{0.727} &0.641&	0.674&	0.697\\

5	&0.767	&0.774&	0.783	&	\textbf{0.828}	&\textbf{0.839}&	\textbf{0.845} &0.802&	0.811&	0.815\\

10&	0.871&	0.878	&0.880&	\textbf{0.910}&	\textbf{0.914}&	\textbf{0.916} &0.880&	0.886&	0.887\\

20&	0.940&	0.944&	0.944	&	\textbf{0.961}&	\textbf{0.962}	&\textbf{0.963} &0.934&	0.936&	0.936\\
50&	0.987&	0.987&	0.987	&	\textbf{0.992}&	\textbf{0.991}&	\textbf{0.991}& 0.969&	0.968	&0.968\\
    \hline
  \end{tabular}
    \end{minipage} 
\end{table*}
\end{scriptsize}

\section{Conclusion and Discussion}
In this work, we provide a thorough investigation of the sampling top-$k$ Hit-Ratio and show how it ``samples'' the global Hit-Ratio in a way similar to ``signal sampling''. Theoretically and empirically, we demonstrate the predictive power of sampling metrics,  in terms of both approximating corresponding global Hit-Ratio, and predicting the relative performance between different algorithms. 

We would like to point out that the mapping function serves as a scaffold for us to understand how sampling works, with respect to the global Hit-Ratio curve. It provides us with a basic tool to help verify the accuracy of the metric being observed from the sampling, with respect to the global curve. Following our theoretical investigation and experimental testing/evaluating of different mapping functions (Section~\ref{experiments}), we can safely use the sampling Hit-Ratio metrics without worrying about the mapping function. 

 %to enable us to establish the validity of the sampling method.

However, there are several interesting open questions which need further investigation.
(1) Can other sampling-based metrics, such as average precision and nDCG, have such properties as the Hit-Ratio (recall/precision)?
(2) Can other distributions fit Hit-Ratio curves better than the beta distribution?
(3) Can the monotone property of the mapping function (with respect to the parameter $a$ in beta distribution) hold beyond Beta distribution?
We plan to (and also welcome the community to) investigate these questions. 

\bibliographystyle{ACM-Reference-Format}
\balance
\bibliography{bib/sample-base}

\section{Acknowledgments}
The research was partially supported by a sponsorship research agreement between Kent State University and iLambda, Inc.

\newpage
\appendix
\section{Experiments and Reproducibility}
\label{ap:ex}

\noindent{\bf Experimental Setup:}
We use four of the most commonly used datasets for recommendation studies, together with the {\texttt book-X} dataset, with a relatively large number of total items, for evaluating the effect of sampling size. The characteristics of these datasets are in Table~\ref{tab:datasets}. 

For the different recommendation algorithms, we use some of the most well-known and the state-of-the-art algorithms, including three non-deep-learning options: itemKNN~\cite{DeshpandeK@itemKNN};
ALS~\cite{hu2008collaborative}; and EASE~\cite{Steck_2019}; 
and three deep learning options:   
NCF~\cite{he2017neural};
MultiVAE~\cite{liang2018variational}; 
and NGCF~\cite{Wang_2019}. 
For each recommendation algorithm on a particular dataset, we report the result of only one sample run, as we found they are very close to the average of multiple run results. The default sampling method is sampling with replacement, unless explicitly stated. Also, we collect both sampling hit ratio $SHR@k$ ($k$ from $1$ to $n$, the sampling size) and global hit ratio $HR@K$ ($K$ from $1$ to $N$, the number of total items). In the figures, we also use {\em population} for $HR@K$ (global) and {\em sample} for $SHR@k$ curves. 
For the different mapping functions, we consider the linear (Formula~\ref{eq:fk1}), the bound (Formula~\ref{eq:bound_mid}), $Beta@1$ ($f(k;a=1)$, Formula~\ref{eq:uni_fk}), $Beta@0.5$ ($f(k;a=0.5)$, Formula~\ref{eq:fk}), and $Beta@P$ for the algorithm-specific mapping (Formula~\ref{recursive}). 
Below, we will report our experimental findings for the aforementioned questions 2 and 3 in Section~\ref{experiments}. 

\begin{figure}[th]
    \includegraphics[width=\linewidth, scale = 0.9]{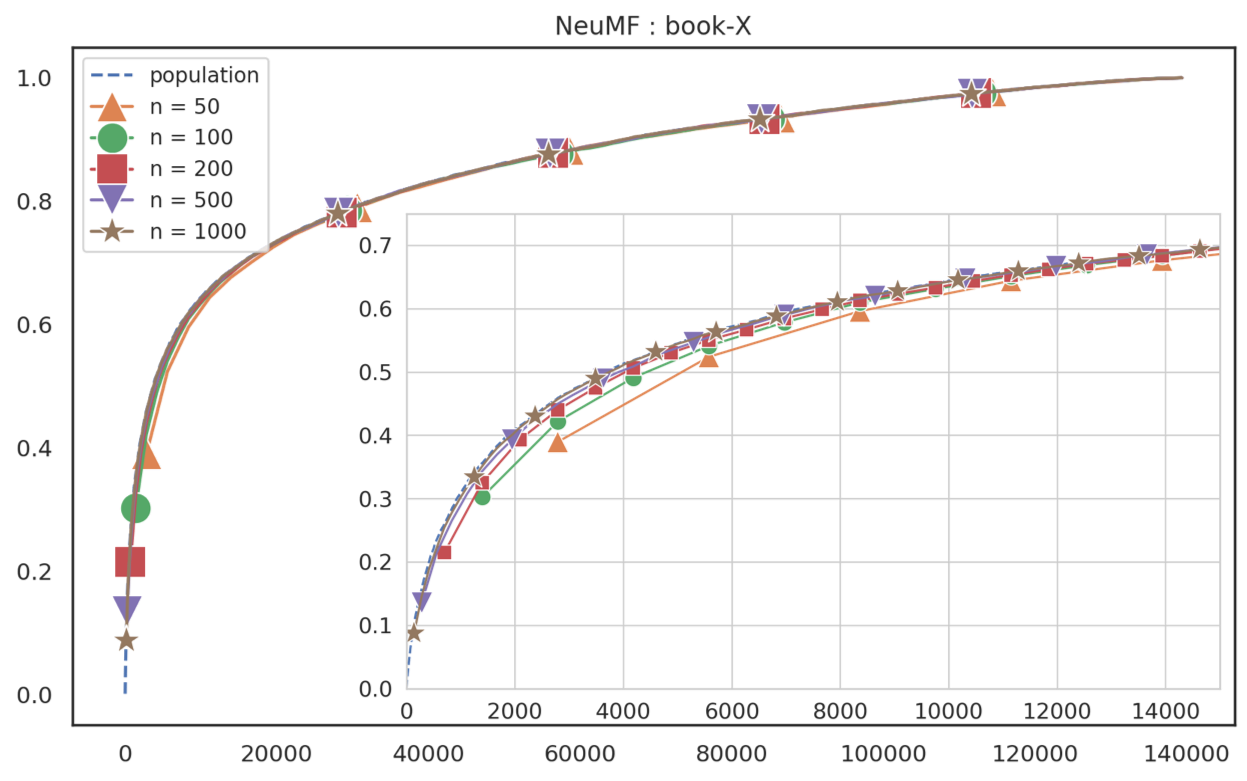}
\caption{Global/Population $HR$ curves compared with sample ones with different sample size. Experiment set-up:dataset:book-X, model:NeuMF}
\label{fig:book_X}
\end{figure}

%----------Question 
\noindent{\bf Key Factor Analysis (Top $k$, Sampling Factors and Sampling Size):}
To take a close look at different mapping functions,  
 we listed the sampling hit ratio $SHR$ at $k=1,2,5,10,20,50$ for sampling size $n-1=99$  and their corresponding global hit ratio $HR$ at $f(k)$ locations based on mapping functions, bound, $beta@1$ and $beta@0.5$ in Table\ref{table:varyingk1}.
 We show similar results for sampling size $n-1=999$. 
 In addition, we also consider three different sampling schemes: sampling with replacement (binom); sampling  without replacement (hyper); and sampling without replacement using only irrelevant items (actual).
 We made the following observations. (1) The differences between the sampling and the global $|SHR@k-HR@f(k)$ are fairly small, with the bound and $Beta@0.5$ more accurate; (2) When $k$ becomes larger, the results are more accurate.  (3) The results on the sampling with replacement and without replacement are very close to each other. Sampling with only irrelevant items, and the Global, which ranks only the irrelevant items, both lead to higher hit ratios. But the mapping function works equally well for this situation. (4) When sampling size increases (from $n=100$ to $n=1000$), the error also reduces. We further confirm this using Figure~\ref{fig:book_X}, which varies $n$ from $50$ to $1000$ on {\texttt book-X} dataset with more $139K$ items. When $n$ increases, the sampling hit ratio curve converges to the global hit ratio (population) rather quickly. 
 
 \noindent{\bf Dataset-independent vs algorithm-specific mapping:} Table~\ref{tab:MappingCompare} compares different dataset-independent mapping functions with the algorithm-specific one, $Beta@P$.  We compare both overall average absolute error $(\sum_{k=1}^n |HR@f(k)-SHR@k|)/n)$, the overall average relative error $(\sum_{k=1}^n (|HR@f(k)-SHR@k|/SHR@k)/n$, the errors at the top $1$, i.e., $|HR@f(1)-SHR@1|$ and $|HR@f(1)-SHR@1|/SHR@1$, and the errors between the top $2$ and $10$.  
 We observe that the algorithm-specific mapping function $Beta@P$ achieves the most minimal errors, over all.  However, the dataset-independent measures, such $Beta@0.5$, obtain comparable results and perform better when $k$ is small. We believe one of the underlying reasons is that $Beta@P$ aims to fit all the users (including those on the lower rank), and the fitting of Beta distribution has limitations. Thus, we consider that the dataset-independent mapping functions, such as $Beta@0.5$, can be a relatively cost-effective way to align sampling and global top-$k$ hit ratio curves. 
 
 % ---- Table

\begin{small}
\begin{table}
  \caption{Dataset Statistics}
  \label{tab:datasets}
  \begin{tabular}{lcccc}
    \toprule
    \textbf{Dataset}&
    \textbf{Interactions}&
    \textbf{Users}&
    \textbf{Items}&\textbf{Sparsity}\\
    \midrule
    ml-1m  &1,000,209 & 6,040 &3,706&95.53$\%$\\
    pinterest-20& 1,463,581 & 55,187&9,916&99.73$\%$\\
    citeulike&204,986 &5,551&16,980&99.78$\%$ \\
    yelp& 696,865&25,677 &25,815&99.89$\%$ \\
    book-X& 786,690&11,325 &139,331&99.95$\%$\\
    \bottomrule
  \end{tabular}
\end{table}
\end{small}

\begin{small}
\begin{table*}[th]
    \caption{Varying Sampling Top $k$  on MultiVAE:ml-1m:n=100}
    \label{table:varyingk1}
    \begin{minipage}{0.32\textwidth}
      \caption*{binom(with replacement)}
      \centering
    \begin{tabular}{|l|cccc|}
    \hline
    \textbf{k}&
    \textbf{SHR}&
    \textbf{bound}&
    \textbf{beta@1}&
    \textbf{beta@0.5}\\
    \hline
1&0.1031&0.0442&0.1116&0.0829\\
2&0.2041&0.1700&0.2185&0.2012\\
5&0.4157&0.4050&0.4334&0.4209\\
10&0.6200&0.6204&0.6323&0.6258\\
20&0.7992&0.8012&0.8050&0.8028\\
50&0.9651&0.9682&0.9684&0.9682\\
    \hline
  \end{tabular}
    \end{minipage}
    \begin{minipage}{0.32\textwidth}
      \centering
        \caption*{hyper(without replacement)}
        \begin{tabular}{|l|cccc|}
    \hline
    \textbf{k}&
    \textbf{SHR}&
    \textbf{bound}&
    \textbf{beta@1}&
    \textbf{beta@0.5}\\
    \hline
1&  0.1012& 0.0442&	0.1116&	0.0829\\
2&	0.1964&	0.1700&	0.2185&	0.2012\\
5&	0.4119&	0.4050&	0.4334&	0.4209\\
10&	0.6157&	0.6204&	0.6323&	0.6258\\
20&	0.7975&	0.8012&	0.8050&	0.8028\\
50&	0.9661&	0.9682&	0.9684&	0.9682\\
    \hline
  \end{tabular}
    \end{minipage}
    \begin{minipage}{0.32\textwidth}
      \centering
        \caption*{actual(training data rank bottom)}
        \begin{tabular}{|l|cccc|}
    \hline
    \textbf{k}&
    \textbf{SHR}&
    \textbf{bound}&
    \textbf{beta@1}&
    \textbf{beta@0.5}\\
    \hline
1&	0.2101&	0.1608&	0.2540&	0.2161\\
2&	0.3349&	0.3199&	0.3810&	0.3555\\
5&	0.5533&	0.5603&	0.5889&	0.5768\\
10&	0.7164&	0.7295&	0.7391&	0.7334\\
20&	0.8531&	0.8661&	0.8690&	0.8674\\
50&	0.9757&	0.9800&	0.9803&	0.9798\\
    \hline
  \end{tabular}
    \end{minipage}
\end{table*}
\end{small}
%---------ml-1m,n=1000------
\begin{small}
\begin{table*}[th]
    \caption{Varying Sampling Top $k$ on MultiVAE:ml-1m:n=1000}
        \label{table:varyingk2}
    \begin{minipage}{0.32\textwidth}
      \caption*{binom(with replacement)}
      \centering
    \begin{tabular}{|l|cccc|}
    \hline
    \textbf{k}&
    \textbf{SHR}&
    \textbf{bound}&
    \textbf{beta@1}&
    \textbf{beta@0.5}\\
    \hline
10&	0.1091&	0.0998&	0.1116&	0.1089\\
20&	0.2126&	0.2124&	0.2185&	0.2174\\
50&	0.4281&	0.4306&	0.4334&	0.4318\\
100&	0.6293&	0.6303&	0.6316&	0.6316\\
200&	0.8036&	0.8043&	0.8050&	0.8046\\
500&	0.9672&	0.9682&	0.9684&	0.9684\\
    \hline
  \end{tabular}
    \end{minipage}
    \begin{minipage}{0.32\textwidth}
      \centering
        \caption*{hyper(without replacement)}
        \begin{tabular}{|l|cccc|}
    \hline
    \textbf{k}&
    \textbf{SHR}&
    \textbf{bound}&
    \textbf{beta@1}&
    \textbf{beta@0.5}\\
    \hline
10&	0.1096&	0.0998&	0.1116&	0.1089\\
20&	0.2126&	0.2124&	0.2185&	0.2174\\
50&	0.4290&	0.4306&	0.4334&	0.4318\\
100&	0.6315&	0.6303&	0.6316&	0.6316\\
200&	0.8060&	0.8043&	0.8050&	0.8046\\
500&	0.9677&	0.9682&	0.9684&	0.9684\\
    \hline
  \end{tabular}
    \end{minipage}
    \begin{minipage}{0.32\textwidth}
      \centering
        \caption*{actual(training data rank bottom)}
        \begin{tabular}{|l|cccc|}
    \hline
    \textbf{k}&
    \textbf{SHR}&
    \textbf{bound}&
    \textbf{beta@1}&
    \textbf{beta@0.5}\\
    \hline
10&	0.2417&	0.2407&	0.2540&	0.2495\\
20&	0.3682&	0.3704&	0.3810&	0.3770\\
50&	0.5776&	0.5861&	0.5889&	0.5874\\
100&	0.7286&	0.7381&	0.7389&	0.7389\\
200&	0.8570&	0.8684&	0.8690&	0.8689\\
500&	0.9747&	0.9800&	0.9803&	0.9803\\
    \hline
  \end{tabular}
    \end{minipage}
\end{table*}
\end{small}

\begin{small}
\begin{table*}[th]
    \caption{Mapping Functions Comparison}
    \label{tab:MappingCompare}
    %\vspace*{-3.0ex}
    \begin{minipage}{0.48\textwidth}
      \caption*{MultiVAE:ml-1m}
      \centering
    \begin{tabular}{|l|cccccc|}
    \hline
    \textbf{Functions}&
    \textbf{abs}&
    \textbf{rel}&
    \textbf{abs@1}&
    \textbf{rel@1}&
    \textbf{abs@2-10}&
    \textbf{rel@2-10}\\
    \hline
Linear&0.0058&0.0226&0.0992&0.9934&0.0391&0.1238\\
Bound&0.0024&0.0103&0.0596&0.5970&0.0100&0.0363\\
Beta@1&0.0034&0.0074&0.0118&0.1177&0.0165&0.0440\\
Beta@0.5&0.0019&0.0043&0.0169&0.1692&0.0052&0.0112\\
Beta@0.2&0.0018&0.0065&0.0356&0.3566&0.0060&0.0207 \\
Beta@P&0.0018&0.0052&0.0250&0.2504&0.0052&0.0146 \\
    \hline
  \end{tabular}
    \end{minipage}\hfill
%     \begin{minipage}{0.48\textwidth}
%       \centering
%         \caption*{EASE:ml-1m}
%         \begin{tabular}{|l|cccccc|}
%     \hline
%     \textbf{Dataset}&
%     \textbf{abs}&
%     \textbf{rel}&
%     \textbf{abs@1}&
%     \textbf{rel@1}&
%     \textbf{abs@2-10}&
%     \textbf{rel@2-10}\\
%     \hline
% Linear&0.0056&0.0211&0.1101&0.9666&0.0374&0.1118\\
% Bound&0.0025&0.0084&0.0500&0.4390&0.0092&0.0294\\
% Beta@1&0.0037&0.0081&0.0205&0.1802&0.0152&0.0408\\
% Beta@0.5&0.0019&0.0031&0.0055&0.0480&0.0040&0.0085\\
% Beta@0.2&0.0018&0.0047&0.0237&0.2078&0.0047&0.0137\\ 
% Beta@P&0.0017&0.0034&0.0141&0.1235&0.0027&0.0059\\ 
%     \hline
%   \end{tabular}
%     \end{minipage} \hfill
%     \begin{minipage}{0.48\textwidth}
%       \caption*{itemKNN:yelp}
%   \begin{tabular}{|l|cccccc|}
%     \hline
%     \textbf{Dataset}&
%     \textbf{abs}&
%     \textbf{rel}&
%     \textbf{abs@1}&
%     \textbf{rel@1}&
%     \textbf{abs@2-10}&
%     \textbf{rel@2-10}\\
%     \hline
% Linear&0.0057&0.0145&0.3019&0.9874&0.0240&0.0443\\
% Bound&0.0019&0.0035&0.0529&0.1731&0.0053&0.0086\\
% Beta@1&0.0040&0.0069&0.0612&0.2003&0.0230&0.0384\\
% Beta@0.5&0.0024&0.0036&0.0183&0.0599&0.0131&0.0212\\
% Beta@0.2&0.0015&0.0022&0.0113&0.0369&0.0067&0.0101\\ 
% Beta@P&0.0017&0.0023&0.0023&0.0075&0.0086&0.0135\\ 
%     \hline
%   \end{tabular}
%     \end{minipage}\hfill
    \begin{minipage}{0.48\textwidth}
      \centering
        \caption*{NeuMF:yelp}
  \begin{tabular}{|l|cccccc|}
    \hline
    \textbf{Functions}&
    \textbf{abs}&
    \textbf{rel}&
    \textbf{abs@1}&
    \textbf{rel@1}&
    \textbf{abs@2-10}&
    \textbf{rel@2-10}\\
    \hline
Linear&0.0057&0.0164&0.2329&0.9886&0.0300&0.0632\\
Bound&0.0019&0.0046&0.0606&0.2571&0.0061&0.0125\\
Beta@1&0.0038&0.0070&0.0414&0.1756&0.0221&0.0413\\
Beta@0.5&0.0020&0.0030&0.0032&0.0134&0.0112&0.0197\\

Beta@0.2&0.0015&0.0027&0.0253&0.1075&0.0057&0.0097 \\
Beta@P&0.0015&0.0024&0.0145&0.0617&0.0067&0.0106\\
    \hline
  \end{tabular}
    \end{minipage} 
    %\vspace*{-3.0ex}
\end{table*}
\end{small}

\section{EM}
\label{sec:em}

\subsection{A little bit review}
\noindent{\bf Considering sampling with replacement.}
For a single user $u$, one time Bernoulli $X_i^u$ is given by:
\begin{small}
\begin{equation*}\label{eq:1}
    X_i^u \sim Bernoulli(p_u = \frac{R_u - 1}{N-1})
\end{equation*}
\end{small}
Random variable $X^u$ denotes the number of sampled items that are ranked in front of relevant item $i_u$ in total $n-1$ samples. And $X^u$ follows binomial distribution: 
\begin{small}
\begin{equation*}\label{eq:2}
\begin{split}
    &X^u = \sum\limits_{i=1}^{n-1}{X_i^u}\\
    &X^u \sim Binomial(n-1, p_u=\frac{R_u - 1}{N-1}) 
    \end{split}
\end{equation*}
\end{small}
Random variable $r^u$ is defined as $r^u = X^u + 1$, denotes the relevant item rank position. Our basic assumption is : $R_u$, denoting the rank position of relevant item among total items, satisfies the distribution $W_R$:
\begin{small}
\begin{equation*}\label{eq:3}
\begin{split}
&r^u \sim W_r,\quad Pr(r^u = r) = W_r,\\
&R_u \sim W_R,\quad Pr(R_u = R) = W_R,
\quad W_R = a(\frac{R-1}{N-1})^{a-1}\frac{1}{N-1}
\end{split}
\end{equation*}
\end{small}
Thus, we have:

\begin{small}
\begin{equation*}\label{eq:4}
\begin{split}
& Pr(r^u = r|R_u = R) = {\binom{n-1}{r-1}}{p_R}^{r-1}(1-p_R)^{n-r}\\
& Pr(r^u = r, R_u = R) = Pr(r^u = r|R_u = R) \cdot W_{R}\\
& Pr(r^u = r) = \sum\limits_{R = 1}^{N}{{\binom{n-1}{r-1}}{p_R}^{r-1}(1-p_R)^{n-r}\cdot W_{R}}
\end{split}
\end{equation*}
\end{small}
where, $p_R = \frac{R-1}{N-1}$. Simply, we can re-write above equation:
\begin{small}
\begin{equation*}\label{eq:5}
\begin{split}
W_r = Pr(r) = \sum\limits_{R}{Pr(r, R)} = \sum\limits_{R}{Pr(r|R)\cdot W_R}
\end{split}
\end{equation*}
\end{small}
\subsection{Expectation}
Recall equation ~\ref{eq:lr0}

\begin{small}
\begin{equation*}
\begin{split}
&\sum\limits_{R = 1}^{N}{W_R \cdot {\bf 1}_{R\le f(k)}}= \sum\limits_{R = 1}^{N}{W_R\cdot Pr(r^R\le k)}\\
&\sum\limits_{R = 1}^{\lfloor f(k)\rfloor}{W_R}= \sum\limits_{R = 1}^{N}{W_R\cdot Pr(r\le k|R )}
\end{split}
\end{equation*}
\end{small}

Note $f$ is approximate-linear, the integers between $\lfloor f(k -1)\rfloor + 1$ and $\lfloor f(k)\rfloor$ are almost constant (independent with $k$), denoting as $C$.
\begin{small}
\begin{equation*}
\begin{split}
&\sum\limits_{R = \lfloor f(k -1)\rfloor + 1}^{\lfloor f(k)\rfloor}{W_R}= \sum\limits_{R = 1}^{N}{W_R\cdot Pr(r = k|R )}\\
\end{split}
\end{equation*}
\end{small} 
\begin{small}
\begin{equation*}
\begin{split}
C\cdot {W_{R = \widehat{f(k)}}}& \approx \sum\limits_{R = 1}^{N}{W_R\cdot Pr(r = k|R )}\\
& = \E\limits_{R\sim W_R}{Pr(r = k|R)}\\
& = Pr(r = k)
\end{split}
\end{equation*}
\end{small}

\subsection{EM}

Define our likelihood function as :
\begin{small}
\begin{equation*}\label{eq:6}
\begin{split}
\mathcal{L}(\theta) = \prod\limits_{u=1}^{M}{Pr(r^u|\theta)}
\end{split}
\end{equation*}
\end{small}
In our EM algorithm, $R$ is the hidden variable.
And the $E$ step is:
\begin{small}
\begin{equation*}\label{eq:7}
\begin{split}
\mathcal{Q}(\theta|\theta^{i}) &=\prod\limits_{u=1}^{M}{Pr(r^u | \theta)}\\
&= \prod\limits_{u=1}^{M}{C\cdot Pr(R_u = f(r^u , \theta^{i})  | \theta)}
\end{split}
\end{equation*}
\end{small}

% \begin{small}
% \begin{equation*}\label{eq:7}
% \begin{split}
% \mathcal{Q}(\theta, \theta^{i}) = \sum\limits_{R = 1}^{N}{\prod\limits_{u=1}^{M}{\Big[Pr(r^u, R|\theta)\Big]}Pr(R|\theta^{i})}
% \end{split}
% \end{equation*}
% \end{small}

And the $M$ step is:
\begin{small}
\begin{equation*}\label{eq:8}
\begin{split}
\theta^{i+1} = arg\max\limits_{\theta} \mathcal{Q}(\theta| \theta^{i})
\end{split}
\end{equation*}
\end{small}

\end{document}